\documentclass[pdflatex,sn-basic,iicol]{sn-jnl}

\usepackage{cancel,ulem,comment}
\usepackage{graphicx}%
\usepackage{multirow}%
\usepackage{amsmath,amssymb,amsfonts}%
\usepackage{amsthm}%
\usepackage{mathrsfs}%
\usepackage[title]{appendix}%
\usepackage{xcolor}%
\usepackage{textcomp}%
\usepackage{manyfoot}%
\usepackage{booktabs}%
\usepackage{algorithm}%
\usepackage{algorithmicx}%
\usepackage{algpseudocode}%
\usepackage{listings}%

\theoremstyle{thmstyleone}%
\newtheorem{theorem}{Theorem}
\theoremstyle{thmstyletwo}%
\newtheorem{remark}{Remark}%
\newtheorem{lemma}{Lemma}%

\theoremstyle{thmstylethree}%

\newcommand{\eps}{\varepsilon}

\begin{document}

\title[Article Title]{Bifurcation of spiking oscillations from a center in
 resonate-and-fire neurons}


\author*[1]{\fnm{Oleg} \sur{Makarenkov}}\email{makarenkov@utdallas.edu}
\equalcont{These authors contributed equally to this work.}

\author[2]{\fnm{Marianne} \sur{Bezaire}}\email{mbezaire@bu.edu}
\equalcont{These authors contributed equally to this work.}

\author[2]{\fnm{Michael} \sur{Hasselmo}}\email{hasselmo@gmail.com}
\equalcont{These authors contributed equally to this work.}

\affil*[1]{\orgdiv{Department of Mathematical Sciences}, \orgname{The University of Texas at Dallas}, \orgaddress{\street{800 West Campbell Road}, \city{Richardson}, \postcode{75080}, \state{Texas}, \country{United States}}}

\affil[2]{\orgdiv{Kilichand Center for Integrated Life Sciences and Engineering}, \orgname{Boston University}, \orgaddress{\street{610 Commonwealth Ave}, \city{Boston}, \postcode{02215}, \state{Massachusetts}, \country{United States}}}


\abstract{The theta rhythm is important for many cognitive functions including spatial
 processing, memory encoding, and memory recall. The information processing underlying
 these functions is thought to rely on consistent, phase-specific spiking throughout a
 theta oscillation that may fluctuate significantly in baseline (center of oscillations), frequency, or amplitude.
 Experimental evidence shows that spikes can occur at specific phases even when the baseline
 membrane potential varies significantly, such that the integrity of phase-locking persists
 across a large variability in spike threshold. The mechanism of this precise spike timing
 during the theta rhythm is not yet known and previous mathematical models have not
 reflected the large variability in threshold potential seen experimentally. Here we introduce a
 straightforward mathematical neural model capable of demonstrating a phase-locked spiking
 in the face of significant baseline membrane potential fluctuation during theta rhythm. This
 novel approach incorporates a degenerate grazing bifurcation of an asymptotically stable oscillation. This model suggests a potential mechanism for how biological neurons can
 consistently produce spikes near the peak of a variable membrane potential oscillation.}




\maketitle

\section{Introduction}
Experimental data indicates that the theta rhythm in hippocampus and associated cortical structures plays an important role in memory function. For example, the efficacy of memory encoding in behavioral tasks correlates with the magnitude of hippocampal theta rhythm \citep{Berry1978, Winson1978, Givens1990}. The theta rhythm, commonly observed as an oscillation in the local field potential, is also visible in individual neuron activity. The spiking of hippocampal neurons in extracellular recordings shows strong dependence on the phase of hippocampal theta rhythm \citep{Fox1986, Buzsaki1983}, and theta rhythmicity is commonly observed in the spiking autocorrelograms of hippocampal \citep{OKeefe1993} and entorhinal neurons \citep{Hafting2008}.

\begin{figure}[h]\center
\vskip-0.2cm
\noindent\includegraphics[scale=1.6]{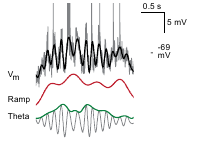}
\vskip-0.0cm
\caption{Experimental observations of grid cell activity, adapted from \citet{Domnisoru2013}. The top trace
shows the intracellular membrane potential (with truncated spikes) in gray. The red trace below shows the
fluctuation in the baseline potential, while the variations in frequency and amplitude of the theta oscillation
are apparent in the theta component shown at the bottom (with envelope in green). The sum of the ramp and
theta components is shown in black overlay at top. The flexibility in spike threshold to enable spiking that is
phase-locked to the peak is clear at the positions marked with black annotations. Figure adapted and reprinted
by permission from Nature Publishing Group.} \label{nature-fig}
\end{figure}

Intracellular recording of hippocampal neurons reveals prominent theta rhythm oscillations of the membrane potential during observation of theta rhythm in the extracellular field potential \citep{Fujita1964, Leung1986, Kamondi1998, Harvey2009}.  Comparison of spike times relative to both intracellular membrane potential and local field potential recordings reveal a consistent phase preference of the timing of spikes relative to the intracellular membrane potential. The spiking activity occurs on the rising phase near the peak of each cycle in hippocampal intracellular recordings by \citet{Harvey2009} and \citet{Kamondi1998} as well as in observations in entorhinal cortex by \citet{SchmidtHieber2013}. The phase preference is also apparent in the entorhinal recordings of \citet{Domnisoru2013}, where the intracellular spike phase preference persists across large variations in theta amplitude and baseline voltage (see Fig.~\ref{nature-fig}). The variations in amplitude and baseline of the theta oscillation are sufficient for the intracellular theta rhythm of the cell to reach membrane potentials within the range of the spike threshold prior to peaking; however, the threshold appears dynamic and able to bias the cell to spike towards the peak of the intracellular rhythm even when the peak potential is relatively depolarized compared to other cycles of the rhythm. The ability of spike phase preference to remain consistent across a range of theta cycle amplitudes has not been widely explored in computational models, although there is robust experimental evidence of the concomitant spike threshold variability \citep{Higgs2011, Platkiewicz2011, Tsuno2013, Wester2013, Fontaine2014}  and functional benefits of spike threshold variability have been modeled \citep{Itskov2011, Huang2016}. 

Many neurons show theta phase precession relative to the theta rhythm in the field potential \citep{OKeefe1993, Skaggs1996, Huxter2003, FernandezRuiz2017}, which could be interpreted as instability of firing phase. However, the studies cited above specifically show that spiking maintains phase relationships to intracellular membrane potential, while both spiking and intracellular membrane potential shift in phase relative to the field potential oscillations \citep{Harvey2009, Domnisoru2013, SchmidtHieber2013}. Models of oscillatory interference can produce spikes phase-locked to the peak of the intracellular theta oscillation \citep{Bush2014}, although they generally assume a constant theta baseline and stable amplitude. Notably, a continuous attractor network model also produced spikes that were phase-locked to the intracellular theta peak when combined with an oscillatory interference model, but not when combined with a slow, depolarizing ramp \citep{SchmidtHieber2013}. Other models start from an assumption of phase-locked spiking to study other phenomena rather than examining the factors causing the phase-locked spikes \citep{OKeefe1993, Lengyel2003, OKeefe2005}. 

{\color{black}In the present paper we use a planar neural model of \cite{Izhikevich2001} to demonstrate mathematically how simple intrinsic dynamics within a neuron or in local circuits of neurons can maintain a phase-locked intracellular spike preference robust to changes in theta amplitude. The novelty of our approach is that it not only predicts a range of parameters where the response of Izhikevich's model is phase-locked (with 1 spike per oscillation),  but also generates spikes on the crest of a wave, rather than appearing on the rising phase of a wave, resembling some typical aspects of the dynamics of Fig.~\ref{nature-fig} of \citet{Domnisoru2013}.}

Our mathematical analysis is based on bifurcation theory. We introduce a small parameter $\eps\ge 0$ in such a way that the model admits a family of cycles for $\eps=0$ and one of the cycles just touches the firing threshold. Such a cycle is known as grazing cycle in nonsmooth bifurcation theory, see \citep{Champ} and \citep{MakarenkovLamb}. The configuration where the grazing cycle is embedded into a family of cycles is called {\it degenerate} because it can be destroyed by small perturbation. It has been already noticed in the literature that the degenerate grazing cycle in impulsive systems is capable to create asymptotic stability under certain types of perturbations, see \citep{Xiaopeng,Yagasaki,Turaev}. In the present paper we use a degenerate grazing cycle to prove the occurrence (termed {\it grazing bifurcation}) of asymptotically stable spiking oscillations in a simple neuron model. Various dynamics of spiking neuron models resulting from grazing bifurcations has been discussed in \citep{Coombes,Nicola}, but grazing bifurcation of asymptotically stable oscillations hasn't been addressed in the earlier literature. To catch the required phenomenon we establish a new theorem about bifurcation of stable periodic solutions from a grazing cycle in impulsive systems, which might be of independent interest in nonsmooth bifurcation theory.

\section{Methods}
{\color{black}We used Izhikevich's resonate-and-fire neuron model (\cite{Izhikevich2001,Izhikevich2006}) to prove the existence of an asymptotically stable spiking cycle with some key differences. As in Izhikevich's model, our reset condition is triggered by the voltage trajectory crossing a defined threshold. However, the reset condition in Izhikevich's resonate-and-fire model acts as $h\to 0$, while we consider $h\to h+\Delta h$ in our analysis. The later impact condition allows us to have a controllable estimate of the magnitude of the impact that we use in our perturbation approach. Reset conditions different from the one in \cite{Izhikevich2001,Izhikevich2006} in relation to resonate-and-fire neuron model have been also considered in e.g. \cite{ImpactLaw2} and \cite{ImpactLaw3}.

\vskip0.2cm

}


\vskip0.2cm

\noindent {\color{black} The second key difference is that unlike most of the methods in neuroscience (see e.g. \cite{Izhikevich2006}) that use global information, such as unstable focus (akin to producing spontaneous intracellular activity in a neuron), or Hopf bifurcation, to generate cyclic spiking behavior, our method is based on a novel {\it grazing bifurcation} and uses local information near one particular trajectory (of the associated unperturbed model) touching the reset threshold.   For the  parameters that we consider for the most of the paper, the model under consideration does indeed have an unstable focus, but we also give an example of parameters (Section~\ref{sec:stablefocus}) where our perturbation approach produces spiking oscillations that surround an asymptotically stable focus. Our theorem provides formulas as for how the reset condition in the former case should be different from reset condition in the later case for asymptotically stable spiking oscillations to exist in both cases.} All numerical simulations were conducted using Mathematica 12.1. 

\subsection{An orbitally stable nearly grazing limit cycle}
\begin{figure}[h]\center
\hskip-0.1cm \includegraphics[scale=0.58]{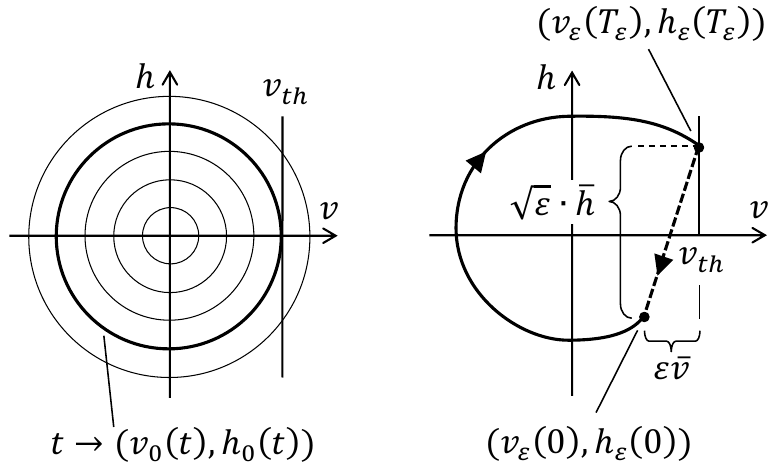}
\caption{\footnotesize Left: Trajectories of system (\ref{reduced}). Right: Discontinuous limit cycle of system (\ref{impulse1})-(\ref{impulse2}) with one impact per period that this paper intends to establish.} \label{iiff}
\end{figure}

First, we determined the conditions for the existence of an orbitally stable nearly grazing limit cycle in a non-linear system with resets. Starting from a set of model equations for a resonate-and-fire model of \cite[\S8.1.2-\S8.1.4]{Izhikevich2006} (see also \cite{Hasselmo2014}), we incorporated a small parameter $\eps$  {\color{black}that destroys the center phase portrait}, as shown in equation (\ref{impulse1}).
\begin{eqnarray}
&& \hskip-1.5cm \begin{array}{rcl}
         \dot v&=&\eps  m_1 v + k_1 h+\eps m  v^2,\\
         \dot h&=&k_2 v +\eps m_2 h,
     \end{array}\label{impulse1}\\
&& \hskip-1.5cm  
\begin{array}{l}v(t)\to v(t)+\eps\bar v,\\ 
h(t)\to h(t)+\sqrt{\eps}\cdot\bar h,
\end{array}\quad{\rm if}\quad v(t)=v_{th},\label{impulse2}
\end{eqnarray}
In equations (\ref{impulse1})-(\ref{impulse2}), $v$ represents the membrane potential of the model neuron in arbitrary units, $t$ represents the time in arbitrary units, and $h$ represents the activation of a hyperpolarization-activated cation channel in arbitrary units. The parameters $m,$ $m_1$, $m_2$, $\bar v$ can be of arbitrary signs including zeros, while parameters $k_1$, $k_2$,  $\bar h,$ $v_{th}$ have to satisfy
\begin{equation}\label{k1k2}
  k_1>0, \  k_2<0,\  v_{th}>0,\  \bar h\not=0,\  0 < \eps\ll 1.
\end{equation}
{\color{black}Note that these constraints correspond to biological properties of the $h$ current. The positive value of $k_1 > 0,$ represents the fact that the h current is a cation current, and the negative value of $k_2 < 0,$ represents the fact that the h current is a hyperpolarization activated current that is more activated when v reaches more hyperpolarized (negative) values. The threshold $v_{th}$ must have a positive value corresponding to a quiescent state of the neuron at resting potential.

The parameters $v_{th}$, $\bar v$, and $\bar h$ define the reset rule.}  This system produces a {\color{black}discontinuous limit cycle} shown in Figure~\ref{iiff} (right).

\vskip0.2cm

\noindent Along with system of equations (\ref{impulse1})-(\ref{impulse2}), consider the reduced system where $\eps = 0$, that gives
\begin{equation}\label{reduced}
 \begin{array}{rcl}
         \dot v&=&k_1 h,\\
         \dot h&=&k_2 v.
     \end{array}
\end{equation}

\noindent Equilibrium $(v,h)=0$ of equation (\ref{reduced}) is of center type and solutions of equation (\ref{reduced}) form a family cycles of period $ T_0=2\pi/\sqrt{-k_1 k_2}$, see Fig.~\ref{iiff} left. {\color{black}Each such cycle is Lyapunov stable, but not asymptotically stable. In what follows, we prove that when $\eps$ changes from $\eps=0$ to small $\eps>0$ the family of cycles  disappears except for the cycle that crosses $v=v_{th}$ transversely (bold cycle in Fig.~\ref{iiff} left). The grazing cycle persists and transforms to an asymptotically stable discontinuous limit cycle of (\ref{impulse1})-(\ref{impulse2}) as $\eps$ crosses 0 (Fig.~\ref{iiff} right). Such a transformation of dynamic behavior (from a Lyapunov stable family of cycles to an asymptotically stable isolated cycle) is usually referred to as {\it bifurcation} in the literature, see e.g. \cite[Chapter II, Definition~2.3]{Chris}.

\vskip0.2cm

\noindent To prove persistence of the grazing cycle as $\eps$ crosses 0, we show that the cross-section $v=v_{th}$ induces a Poincar\'e map $h\mapsto P_\eps(h)$ of system (\ref{impulse1})-(\ref{impulse2}). All points of $(-\infty,0]$ are fixed points for $P_0$. Using the Taylor expansion of $h\mapsto P_\eps(h)$ near $(h,\eps)=(0,0)$, we prove that the fixed point $h=0$ of $P_0$ transforms to an asymptotically stable fixed point $h_\eps$ of $P_\eps$, i.e. $h_\eps\to 0$ as $\eps\to 0$. Moreover, we provide a formula for the limit, as $\eps\to 0$ of the {\it characteristic multiplier} $\rho_\eps=(P_\eps)'(h_\eps)$ of the discontinuous limit cycle corresponding to the fixed point $h_\eps$. Remarkably, the limit of $\rho_\eps$ is different from 1 even though the characteristic multiplier of any cycle of reduced system (\ref{reduced}) is 1, which discontinuity in the behavior of $\rho_\eps$ is typical for grazing bifurcations.} 

\vskip0.2cm

\noindent
The central role in our analysis is played by the following constants $a_0$ and $b_0$:
\begin{equation}
\begin{array}{l}\label{constants}
a_0=\dfrac{1}{k_2 v_{th}},\  b_0=2\dfrac{\dfrac{\pi(m_1+m_2)}{\sqrt{-k_1 k_2}}v_{th}+\bar v}{k_1 k_2 v_{th}},
\end{array}
\end{equation}
which are defined if condition (\ref{k1k2}) holds. The core of our method is the following theorem.

\message{LaTeX Warning: Can Oleg add a transitional sentence that explains the motivation for the following theorem here? \the\inputlineno}

\begin{theorem}\label{neuronthm} 
Assume that system (\ref{reduced}) is of center type, i.e. (\ref{k1k2}) holds.
Denote by 
 $t\mapsto(v_0(t),h_0(t))$ the cycle of (\ref{reduced}) that touches the line $v=v_{th}>0$ (bold cycle in Fig.~\ref{iiff} left). If 
{\color{black}\begin{eqnarray}
&&    \bar h\not=0,\quad  b_0<0, \label{A1} \\ 
&& \dfrac{\bar h^2+b_0/a_0^2}{2\bar h}+\bar h<0, \label{condition2} 
\end{eqnarray}}
then, for each $\eps>0$ sufficiently small, the system (\ref{impulse1})-(\ref{impulse2}) admits an asymptotically stable $T_\eps$-periodic limit cycle 
$t\mapsto(v_\eps(t),h_\eps(t))$ of one impact per period and satisfying 
\begin{eqnarray*}
&& (v_\eps(t),h_\eps(t))\to (v_0(t),h_0(t)), \ T_\eps\to T_0, \ {\rm as}\   \eps\to 0,
\end{eqnarray*}
where $T_0$ is the minimal period of the cycle $(v_0,h_0).$ The characteristic multiplier $\rho_\eps$ of the limit cycle $(v_\eps,h_\eps)$ satisfies   
{\color{black}\begin{equation}\label{stabcond}
    \lim\limits_{\eps\to 0}\rho_\eps=\rho_0,\ \ {\rm where}\ \ \rho_0=\dfrac{\bar h^2+b_0/a_0^2}{\left|\bar h^2-b_0/a_0^2\right|}{\rm sign}(\bar h).
\end{equation}}
\end{theorem}
\noindent We note that condition (\ref{A1}) ensures that $\rho_0\in(-1,1).$

\noindent Throughout the paper, same notation $I(\cdot)$ stays for various (different) functions whose
explicit formula is not used in proofs (and doesn't influence conclusions).

\vskip0.2cm

\noindent {\color{black}We remind the reader that, for a scalar function of 3 variables $(t,h,\eps)\mapsto f(t,h,\eps)$, the Taylor Theorem for power series of order 2 about $(t,h,\eps)=0$ reads as, see e.g. \cite[Ch.~XIII, \S6]{Lang},
$$
\begin{array}{l}
f(y)=f(0)+Df(0)y+\dfrac{D^2f(0)}{2}y^2+\dfrac{D^3f(y_*)}{6}y^3,
\end{array}
$$
where 
$$y=\left(\begin{array}{c} t\\ h\\ \eps\end{array}\right), \quad y_*=\left(\begin{array}{c} t_*\\ h_*\\ \eps_*\end{array}\right),$$
and $y^2$ is 2-tuple $(y,y)$, $y^3$ is 3-tuple $(y,y,y)$,
 $Df$, $D^2f$, $D^3 f$ are first, second, and third Fr\'echet derivatives of $f$ with respect to the vector $(t,h,\eps).$ The term $D^2f$ is called Hessian or tensor of order 2, and $D^3 f$ is tensor of order 3. The space of $2$-tuples of vectors of $\mathbb{R}^3$ is denoted $R^{3\times 3}$ and the space of $3$-tuples of vectors of $\mathbb{R}^3$ is denoted $\mathbb{R}^{3\times 3\times 3}.$} 

\begin{lemma} \label{lemsingular} {\color{black}Let, for every $(t,h,\eps)\in\mathbb{R}^3$, the function $$I(t,h,\eps):\mathbb{R}^{3\times 3\times 3}\to\mathbb{R}$$ be a tensor of order 3, that is $C^2$ smooth with respect to the variables $t,h,\eps.$} Consider the equation \begin{equation}\label{genform}
   t^2+at h+b\eps+ct\eps+p h\eps+ q\eps^2+I(t,h,\eps)\left(\begin{array}{c} t\\ h\\ \eps\end{array}\right)^3=0,
\end{equation}
{\color{black}where $a,b,c,p,h,q\in\mathbb{R}$ are arbitrary constants}.
Then, given $\gamma>0$ there exists $\delta>0$ such that for any $h,\eps\,{\color{black}\not=0}$ that satisfy
\begin{equation}\label{separated}
   \sqrt{h^4+\eps^2}\le \delta\quad{and}\quad \dfrac{(ah)^2-4b\eps}{\sqrt{h^4+\eps^2}}\ge \gamma
\end{equation}
the equation (\ref{genform}) admits exactly two solutions $t$. The least of the two solutions is given by  
$$
  t=\dfrac{1}{2}\left(-ah-\sqrt{(ah)^2-4b\eps}\right)+I(h,\eps)\sqrt{h^4+\eps^2},
$$
where $I$ is $C^1$ in $\sqrt{h^4+\eps^2}\le \delta.$
\end{lemma}

\noindent Even though the requirement of $\eps>0$ is needed for Theorem~\ref{neuronthm}, the proof of Lemma~\ref{lemsingular} works for any $\eps\not=0$, so we didn't restrict the sign of $\eps$ in the formulation of Lemma~\ref{lemsingular}.

\vskip0.2cm

{\color{black}
\noindent To formulate the next lemma we denote by $t\mapsto (V,H)(t,v_0,h_0,\eps)$ the solution $t\mapsto (v,h)(t)$ of system (\ref{impulse1}) with the initial condition $(v,h)(0)=(v_0,h_0).$ 

\vskip0.2cm

\noindent In what follows, for functions of several variables, a letter in subscript refers to partial derivative with the respect to the corresponding variable, for example $V_h$ is partial derivative of $V$ with respect to $h$ and $V_{vh}$ is mixed partial derivative of $V$ with respect to $v$ and $h$. For functions of one variable, a letter in subscript refers to parameter, i.e. a function $P_\eps:\mathbb{R}\to\mathbb{R}$ is an $\eps$-dependent function and $(P_\eps)'$ is the derivative of such a function.

\vskip0.2cm

\noindent When it doesn't cause confusion, vectors rows are identified with vectors columns. For example, in Lemma~\ref{lemma2} below the 4-dimensional vector $x_0=(T,v,h,\eps)$ stays for $(T,v,h,\eps)^T.$}

\begin{lemma}\label{lemma2} {\color{black} Let $T_0>0$ be the minimal period of the cycle $(v_0(t),h_0(t))$ as defined in Theorem~\ref{neuronthm}.} With the notation $x_0=(T_0,v_{th},0,0)$, it holds
$$
\begin{array}{l}
\left(\begin{array}{cc}
    V_v(x_0) & V_h(x_0)\\
    H_v(x_0) & H_h(x_0)\end{array}
\right)=\left(\begin{array}{cc}
  1 & 0 \\ 0 & 1 
    \end{array}
\right),\\ \left(\begin{array}{c}
    V_{\eps}(x_0)\\
    H_{\eps}(x_0)\end{array}
\right)=\dfrac{\pi (m_1+m_2)}{\omega}\left(\begin{array}{c}
  v_{th}\\ 0 
    \end{array}
\right),\quad \omega=\sqrt{-k_1k_2}, 
\end{array}
$$
$$
\begin{array}{l}
V_t(x_0)=0,\quad H_t(x_0)=k_2 v_{th},\quad a_0H_t(x_0)=1,\\
   V_{vv}(x_0)=V_{vh}(x_0)=V_{hh}(x_0)=H_{vv}(x_0)=\\\qquad=H_{vh}(x_0)=H_{hh}(x_0)=0,\\
      V_{th}(x_0)=k_1,\quad V_{tt}(x_0)=k_1k_2v_{th}.
\end{array}
$$
\end{lemma}

\vskip0.2cm

\noindent {\bf Proof of Theorem~\ref{neuronthm}.} {\bf Step 1:} {\it Computing the expansion of the time map $T(h,\eps).$}  Expanding ${\color{black}(T,h,\eps)\to} \, \,V\left(T,v_{th}+\eps \bar v, h,\eps\right)$ in Taylor series about $(T,h,\eps)=(T_0,0,0)$ and denoting $x_0=(T_0,  v_{th}, 0,0)$ one has the following equation for time map $T(h,\eps)$:
\begin{eqnarray*}
0&=&-v_{th}+V\left(T, v_{th}+\eps \bar v, h,\eps\right)=\\
&&=-v_{th}+v_{th}+V_t(x_0)(T-T_0)+\\
&&+V_h(x_0)h+V_v(x_0)\bar v\eps+V_\eps(x_0)\eps+\\
&& +
\dfrac{1}{2}V{}_t{}_t(x_0)(T-T_0)^2+V{}_t{}_h(x_0)(T-T_0)h+\\
&&+V{}_t{}_\eps(x_0)(T-T_0)\eps+V_{tv}(x_0)(T-T_0)\bar v\eps+\\ 
&& +\dfrac{1}{2}V{}_h{}_h(x_0)h^2+V{}_h{}_\eps(x_0)h\eps+\dfrac{1}{2}V{}_\eps{}_\eps(x_0)\eps^2+\\
&&+I(T,h,\eps)\left((T-T_0,h,\eps)^T\right)^3.
\end{eqnarray*}
Using Lemma~\ref{lemma2} to delete zero terms and applying Lemma~\ref{lemsingular} (with $a=a_0$ and $b=b_0$) to solve for $T$, we get
\begin{equation}\label{lemma2formula}
   T(h,\eps)=T_0-a_0h -\sqrt{(a_0h)^2-b_0\eps}\ +\ I(h,\eps)\sqrt{h^4+\eps^2},
\end{equation}
where
$$   a_0=\dfrac{V_{th}(x_0)}{V_{tt}(x_0)},\qquad 
b_0=\dfrac{2\left(V_\eps(x_0)+V_v(x_0)\bar v\right)}{V_{tt}(x_0)}.
$$
Condition (\ref{separated}) of Lemma~\ref{lemsingular} will be verified later.

\vskip0.2cm

\noindent {\bf Step 2:} {\it Expanding the Poincar\'e map $P_\eps(h)=H\left(T(h,\eps), v_{th}+\eps \bar v, h,\eps\right)+\sqrt{\eps}\cdot \bar h$ in powers of $(h,\eps).$} Since
$$
\begin{array}{l}
   H\left(T,v_{th}+\eps \bar v , h,\eps\right)=H_t\left(x_0\right)(T-T_0)+\\
   +
H_h\left(x_0\right)h+
\left(H_\eps\left(x_0\right)+H_v(x_0)\bar v\right)\eps +\\
 + I(T,h,\eps)\left((T-T_0,h,\eps)^T\right)^2,
\end{array}
$$
we use (\ref{lemma2formula}) to get
\begin{equation}\label{odef} 
\begin{array}{l}
  P_\eps(h)=\bar P_\eps(h)\ +\  o(h,\eps),\\ o(h,\eps)=\ I(h,\eps)\left(\left(\sqrt{h^4+\eps^2},h,\eps\right)^T\right)^2,
\end{array}
\end{equation}
where
$$
\begin{array}{l}
   \bar P_\eps(h)=-H_t\left(x_0\right) a_0 h-H_t\left(x_0\right)\sqrt{(a_0h)^2-b_0\eps}+\\
   \qquad\qquad+h+\sqrt{\eps}\cdot\bar h.
\end{array}
$$

\begin{figure}[h]\center
\vskip-0.4cm \includegraphics[scale=0.59]{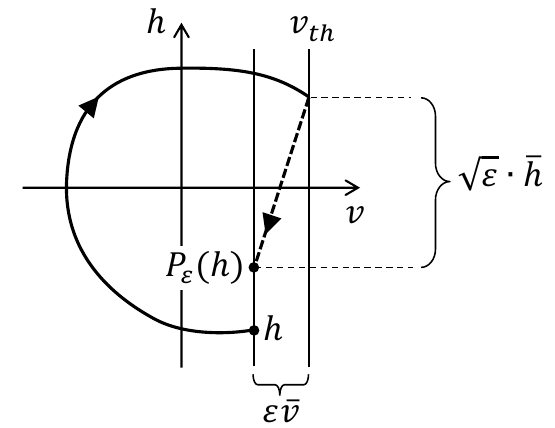}
\caption{\footnotesize Poincar\'e map of system (\ref{impulse1})-(\ref{impulse2}) induced by cross-section $v=v_{th}+\eps \bar v.$} \label{P-map}
\end{figure}

\vskip0.2cm

\noindent{\bf Step 3:} {\it Studying the dynamics of the reduced map $\bar P_\eps(h).$ }
\noindent Solving $\bar P_\eps(h)=h$, one gets $\bar h_\eps=\dfrac{\bar h^2+H_t(x_0))^{\color{black}2}b_0}{2H_t(x_0)a_0\bar h}\sqrt{\eps},$ which can be simplified as 
\begin{equation}\label{barxeps}
\bar h_\eps=\dfrac{\bar h^2+H_t(x_0)^{\color{black}2}b_0}{2\bar h}\sqrt{\eps},
\end{equation}
as $H_t(x_0)a_0=1$ according to formulas of Lemma~\ref{lemma2} and formula~(\ref{constants}). Using same formulas, computation of $(\bar P_\eps)'(\bar h_\eps)$ yields
$$
\begin{array}{l}
(\bar P_\eps)'(\bar h_\eps)=\\
=-{\color{black}H_t(x_0)}\dfrac{(a_0^{\color{black}2}\bar h^2+ b_0)\, {\rm sign}(\bar h)}{\sqrt{(a_0\bar h^2+{\color{black}H_t(x_0)}\,b_0)^2-4\bar h^2b_0}},
\end{array}
$$
which simplifies to $\rho_0$ given by (\ref{stabcond}).

\vskip0.2cm

\noindent {\bf Step 4:} {\it Linking the reduced map $\bar P_\eps$ to the full map $P_\eps$.} First, we show that the map $P_\eps$ has a fixed point $h_\eps$ close to the fixed point $\bar h_\eps$ of the map $\bar P_\eps$ {\color{black}(still leaving verification of condition (\ref{separated}) of Lemma~\ref{lemsingular} for later)}. We will search for $h_\eps$ in the form
\begin{equation}\label{hepsform}    
   h_\eps=\bar h_\eps+\eps^{\color{black}3/4}\tau_\eps,
\end{equation}
{\color{black}where $\tau_\eps$ is an $\eps$-dependent constant to be determined.} Substituting (\ref{hepsform}) to (\ref{odef}) we get the following  equation for $\tau_\eps$:
$$
   \bar P_\eps(\bar h_\eps+\eps^{\color{black}3/4}\tau)+o(\bar h_\eps+\eps^{\color{black}3/4}\tau,\eps)=\bar h_\eps+\eps^{\color{black}3/4}\tau.
$$
Since $\bar P_\eps(\bar h_\eps)=\bar h_\eps$, we obtain
$$
   (\bar P_\eps)'(\bar h_\eps)\eps^{\color{black}3/4}\tau+I(\eps,\tau)(\eps^{\color{black}3/4}\tau)^2+o(\bar h_\eps+\eps^{\color{black}3/4}\tau,\eps)=\eps^{\color{black}3/4}\tau,
$$
and dividing by $\eps^{\color{black}3/4},$ 
$$
   (\bar P_\eps)'(\bar h_\eps)\tau+I(\eps,\tau)\eps^{\color{black}3/4}\tau^2+\dfrac{o(\bar h_\eps+\eps^{\color{black}3/4}\tau,\eps)}{\eps^{\color{black}3/4}}-\tau=0.
$$
Denoting the left-hand-side of the last equation by $F(\tau,\eps)$, we can use the implicit function theorem to solve $F(\tau,\eps)=0$. Indeed, noticing that  (\ref{odef})-(\ref{barxeps}) imply that 
$o(\bar h_\eps+\eps^{\color{black}3/4}\tau,\eps)$ is of order $\eps$, we conclude  
$F(0,0)=0$.
To establish $F_\tau(0,0)\not=0$
we verify that  
$$\dfrac{d}{d\tau}\left(\dfrac{o(\bar h_\eps+\eps^{\color{black}3/4}\tau,\eps)}{\eps^{\color{black}3/4}}\right)=o_h({\bar h}_\eps+\eps^{\color{black}3/4}\tau,\eps)\to 0
$$
as $\eps\to 0$, yielding $F_\tau(0,0)=\rho_0-1\not=0$ by (\ref{A1}). From the implicit function theorem we now obtain $\tau_\eps\to 0$ as $\eps\to 0$ such that $F(\tau_\eps,\eps)=0$ for all $\eps\not=0$ and so (\ref{hepsform}) holds true.

\vskip0.2cm

\noindent To show that stability of the fixed point 
$h_\eps$ of $P_\eps$ coincides with stability of the fixed point $\bar h_\eps$ of $\bar P_\eps$, we observe that 
$$
\begin{array}{l}
\rho_\eps=(P_\eps)'(h_\eps)\to (\bar P_\eps)'(\bar h_\eps)=\rho_0,\quad{\rm as}\ \eps\to 0,
\end{array}
$$
because $
o_h(\bar h_\eps+\eps^{\color{black}3/4}\tau_\eps,\eps)\to 0$ as $\eps\to 0.$
Since (\ref{A1}) implies $\rho_0\in(-1,1)$, we conclude that $\rho_\eps\in(-1,1)$ for $\eps>0$ sufficiently small.

\vskip0.2cm

{\color{black}
\noindent {\bf Step 5:} {\it Verification of condition (\ref{separated}) of Lemma~\ref{lemsingular}.} It is sufficient to show the existence of $\gamma>0$ such that the second inequality of (\ref{separated}) holds with $(h,\eps)=(h_\eps,\eps).$ We have $\dfrac{(2a_0h_\eps)^2-4b_0\eps}{\sqrt{h_\eps^4+\eps^2}}=$
\begin{eqnarray}
&&  =4\dfrac{(a_0\bar h^2+H_t(x_0)b_0+a_0\eps^{1/4}\tau_\eps2\bar h)^2-4b_0\bar h^2}{\sqrt{(\bar h^2+H_t(x_0)^2b_0+\eps^{1/4}\tau_\eps2\bar h)^4+(2\bar h)^4}}\ge\nonumber\\
&&\ge 2a_0^2\dfrac{(\bar h^2+b_0/a_0^2)^2-4b_0\bar h^2/a_0^2}{\sqrt{(\bar h^2+H_t(x_0)^2b_0)^4+(2\bar h)^4}},\label{lastin}
\end{eqnarray}
for all $\eps>0$ sufficiently small. And strict positivity of (\ref{lastin}) follows by assumption (\ref{A1}).
}

\vskip0.2cm

\noindent Let now $t\mapsto (v_\eps(t),h_\eps(t))$ be the solution of (\ref{impulse1})-(\ref{impulse2}) with the initial condition $(v_{th}+\eps\bar v,h_\eps)$, i.e. $t\mapsto (v_\eps(t),h_\eps(t))$ takes the value $(v_{th}+\eps \bar v,h_\eps)$ right after the reset law (\ref{impulse2}) is applied, that we formulate as $(v_\eps(0^+),h_\eps(0^+))=(v_{th}+\eps\bar v,h_\eps).$ Accordingly $t\mapsto (v_\eps(t),h_\eps(t))$ takes the value $(v_{th} ,h_\eps-\sqrt{\eps}\cdot\bar h)$ right before  (\ref{impulse2}) is applied, that is recorded as $(v_\eps(0^-),h_\eps(0^-))=(v_{th},h_\eps-\sqrt{\eps}\cdot\bar h).$

\vskip0.2cm

\noindent {\bf Step 6:} {\it Verifying that  solution $(v_\eps(t),h_\eps(t))$ with the initial condition $(v_{th}+\eps \bar v,h_\eps)$ doesn't cross the firing threshold $v=v_{th}$ on the open interval $(0,T(h_\eps,\eps))$.}  {\color{black} From equation (\ref{impulse1}) we conclude that the set of all points of the vector field of (\ref{impulse1})-(\ref{impulse2}) where $\dot v =0$ is given by 
$$
   h=L(v)\quad {\rm with}\quad L(v)=-{(\eps m_1 v+\eps m v^2)}/{k_1}.
$$

\noindent {\it Claim:} If $v_\eps(t_\eps)=v_{th}$ for $t_\eps\in(0,T(h_\eps,\eps))$ then $\dot v_\eps(s_\eps)=0$ with $h_\eps(s_\eps)\not=L(v_\eps(s_\eps))$ at some $s_\eps\in (0,t_\eps)$ contradicting  definition of the curve $L.$

\begin{figure}[h]\center
\vskip-0.0cm \includegraphics[scale=0.59]{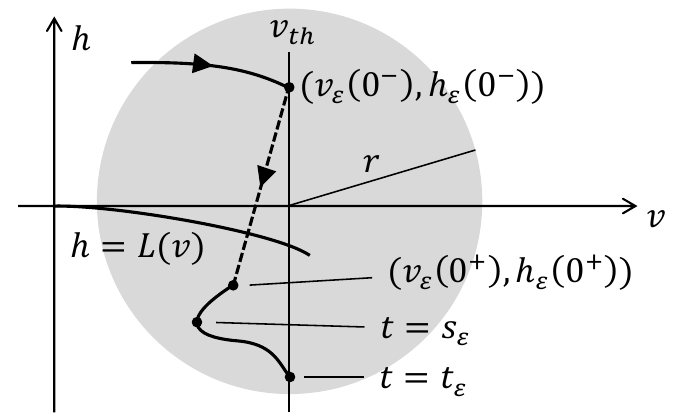}
\caption{\footnotesize Illustration of the proof of the claim.} \label{1impact}
\end{figure}

\noindent {\it Proof of the Claim. Step I.} To prove that  
$$
  h_\eps(0^+)<L(v),\quad\mbox{for all }v\in[0,v_{th}],
$$
we divide this inequality by  $\sqrt{\eps}$ and use (\ref{barxeps})-(\ref{hepsform}) to get the following equivalent   inequality
$$
\begin{array}{l}
  \dfrac{\bar h^2+H_t(x_0)^{\color{black}2}b_0}{2\bar h}+\eps^{1/4}\tau_\eps+\bar h<-\dfrac{\sqrt{\eps}(m_1 v+  m v^2)}{k_1},
\end{array}
$$
which holds for 
all $v\in[0,v_{th}]$ by assumption (\ref{condition2}), if $\eps>0$ is 
sufficiently small. 
\vskip0.2cm

\noindent {\it Step II.} Observing from (\ref{impulse1}) that 
$$
   \dfrac{\dot v_\eps(0^+)}{\sqrt{\eps}}\to k_1\left(\dfrac{\bar h^2+H_t(x_0)^2b_0}{2\bar h}+\bar h\right)\ \ {\rm as}\ \ \eps\to 0,
$$
we use assumptions (\ref{k1k2}) and (\ref{condition2})  to conclude $\dot v_\eps(0^+)<0$ for all $\eps>0$ sufficiently small. Therefore, $v_\eps(t_\eps)=v_{th}$ implies the existence of $s_\eps\in(0,t_\eps)$ such that $\dot v_\eps(s_\eps)=0.$

\vskip0.2cm

\noindent {\it Step III.} Now we fix $r\in(0,v_{th})$ and observe that, when $\eps>0$ is sufficiently small, the vector field (\ref{impulse1})-(\ref{impulse2}) satisfies $\dot h<0$ for any $(v,h)$ from an $r$-neighborhood of $(v_{th},0)$ by assumption (\ref{k1k2}). Therefore, because $h_\eps(0^+)$ is already below $L([0,v_{th}]),$ $h_\eps(s_\eps)$ must be below $L([0,v_{th}])$ as well, i.e. $h_\eps(s_\eps)\not=L(v_\eps(s_\eps))$ and the desired Claim has been achieved.
}

\vskip0.2cm

\noindent

\noindent The proof of the theorem is complete.

\begin{remark} \label{remark1} One can see from the proof of Theorem~\ref{neuronthm} (and also by looking at formula (\ref{stabcond}) for $\rho_0$) that out of two conditions (\ref{A1}) and (\ref{condition2}), condition (\ref{A1}) is responsible for stability of a fixed point of the Poincare map of system (\ref{impulse1})-(\ref{impulse2}) while condition (\ref{condition2}) prohibits repeated impacts of the corresponding periodic cycle (i.e. ensures that exactly 1 spike occurs during 1 period). 
\end{remark}

\subsection{Use of random processes to produce fluctuations in spike threshold}\label{sec22}
To investigate robustness of qualitative predictions of Theorem~\ref{neuronthm}, we incorporate a random process into the threshold condition of our system as follows:
\begin{equation}\label{impulse2p}
\begin{array}{l}
v(t)\to v(t)\,{\color{black}+}\, \eps\overline v,\quad h(t)\to h(t)\,{\color{black}+}\,\sqrt{\eps}\cdot\bar h,\\{\rm if}\quad v(t)=v_{th} + noise(t),
\end{array}
\end{equation}
with $noise(t)$ being a Wiener process, which replaces the deterministic reset condition (\ref{impulse2}).
The addition of a Wiener process provides a slowly varying, stochastic noise term to the threshold condition, allowing for realistic variability in the spike threshold of the system. 

\vskip0.2cm

\noindent We are interested to determine how the phase-locking prediction of Theorem~\ref{neuronthm} (one spike per oscillation) is affected by the noise addition. To do this, we simulate the dynamics of the system of (\ref{impulse1}) and (\ref{impulse2p}) for a range of the   values of $v_{th}$ that satisfy or violate conditions of Theorem~\ref{neuronthm}.

\begin{figure}[h]\center
\hskip0.2cm\noindent\includegraphics[scale=0.5]{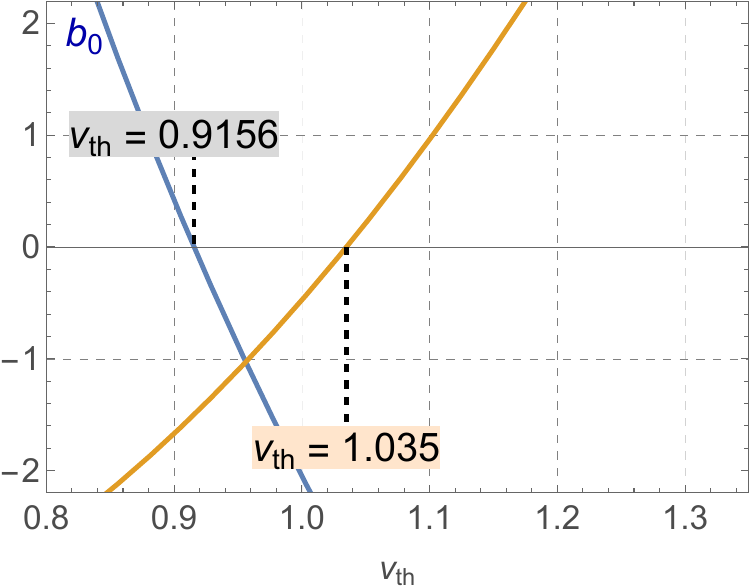}
\vskip-4.0cm\hskip3.9cm {\color{orange}$\dfrac{\bar h^2+b_0/a_0^2}{2\bar h}+\bar h$}
\vskip3.4cm
\caption{Graphs of $b_0$ and $\dfrac{\bar h^2+b_0/a_0^2}{2\bar h}+\bar h$ from (\ref{A1})-(\ref{condition2}) as function of $v_{th}$ with other parameters given by (\ref{param}).} \label{figabrho}
\end{figure}

\section{Results}
\subsection{Autonomous case}\label{simu}

In this section we pick parameters satisfying Theorem~\ref{neuronthm} except for $v_{th}$ that we vary in order to demonstrate the agreement of stability prediction (\ref{stabcond}) with numerical simulations.
 Specifically, considering
\begin{equation}\label{param}
\begin{array}{l}
 k_1=0.18,\ \  k_2=-1,\ \  m_1=0.645,\\  m_2=-0.35,\ \ m=0.3,\ \ \bar h=-1,\ \ \bar v=-2,
\end{array}
\end{equation}
{\color{black}we compute the quantities in (\ref{A1}) and (\ref{condition2}) as functions of $v_{th}.$
The graphs of these functions are shown in  
Fig.~\ref{figabrho}, from where we conclude that, for parameters (\ref{param}) and focusing on $v_{th}>0$, conditions (\ref{A1})-(\ref{condition2}) for the existence of an asymptotically stable 1-period discontinuous limit cycle hold when
\begin{equation}\label{interv1}
      0.9156<v_{th}<1.035.
\end{equation}

\begin{figure*}[!ht]\center
\vskip0.1cm

\noindent\includegraphics[scale=0.38]{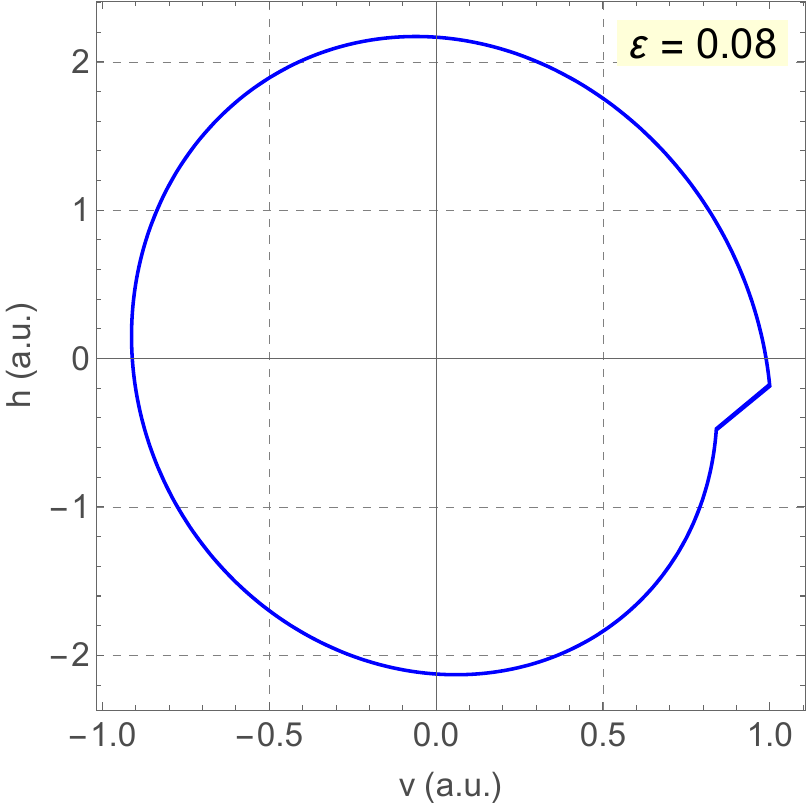} \includegraphics[scale=0.38]{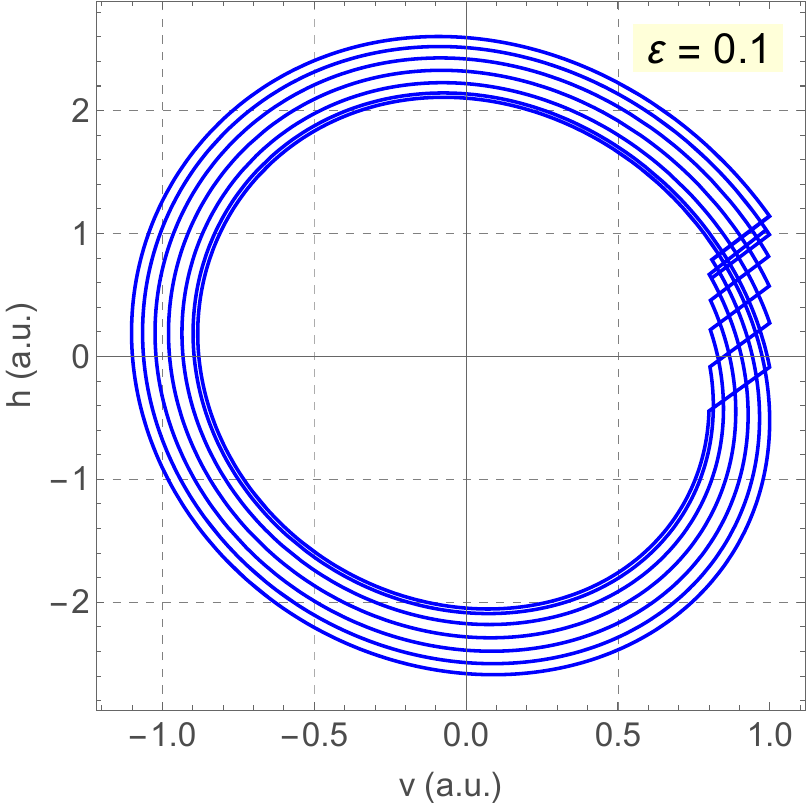} \includegraphics[scale=0.38]{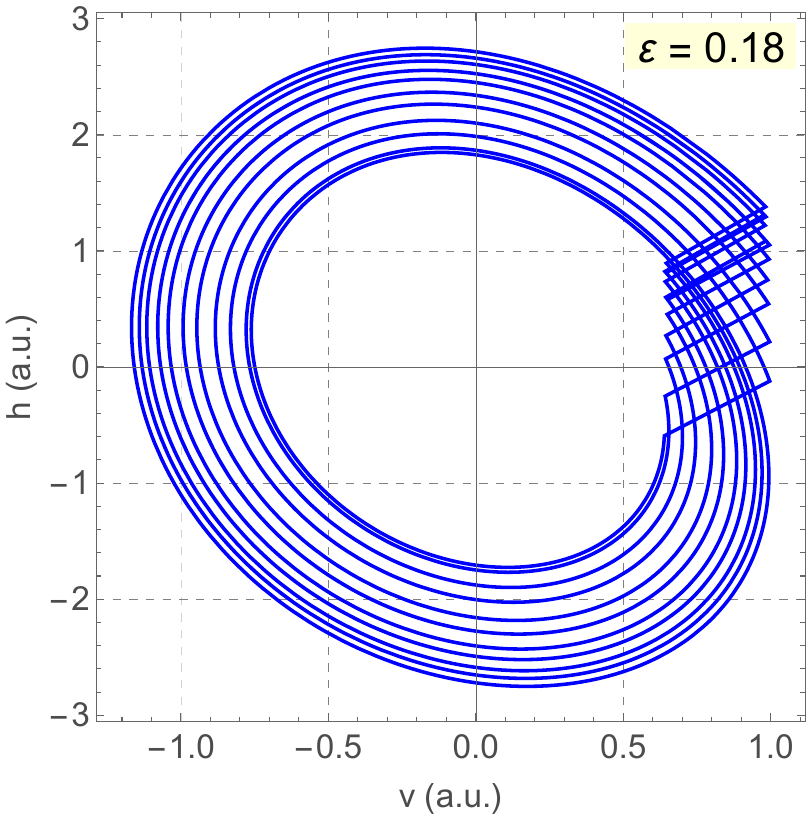}

\vskip0.1cm
\noindent\includegraphics[scale=0.38]{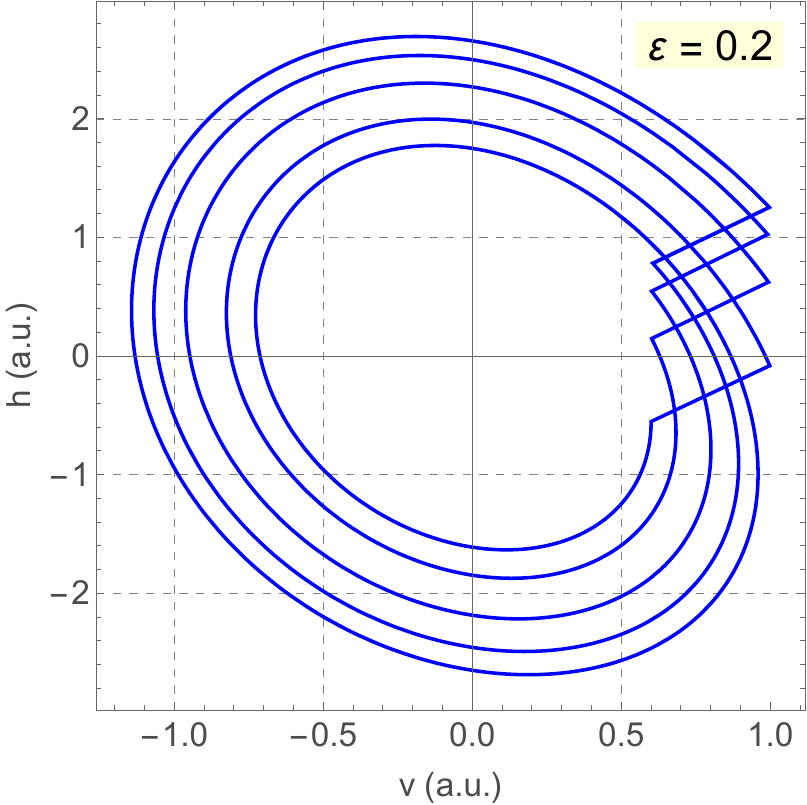}
\includegraphics[scale=0.38]{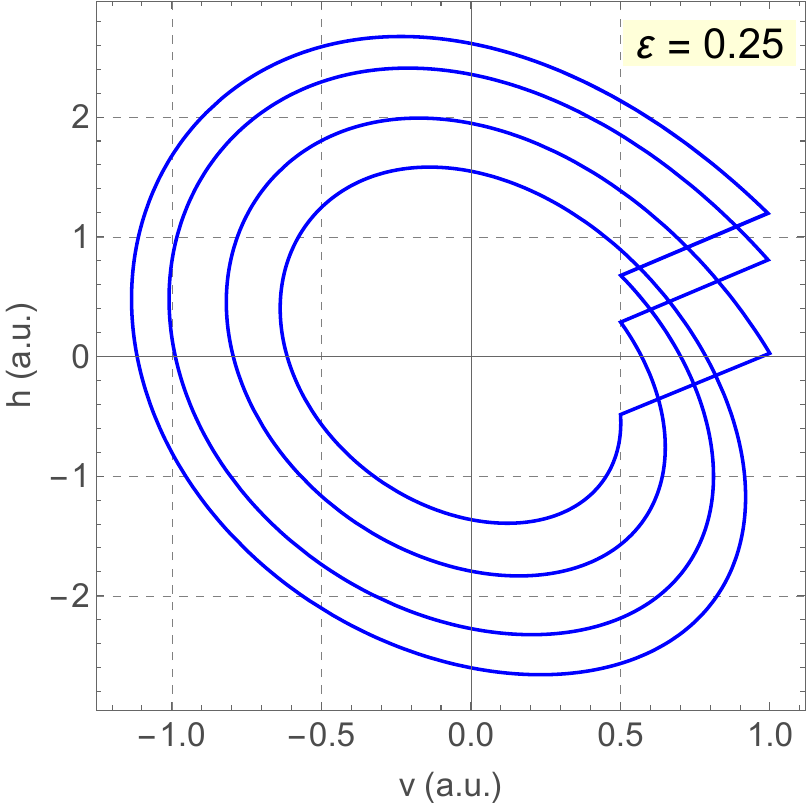}
\includegraphics[scale=0.38]{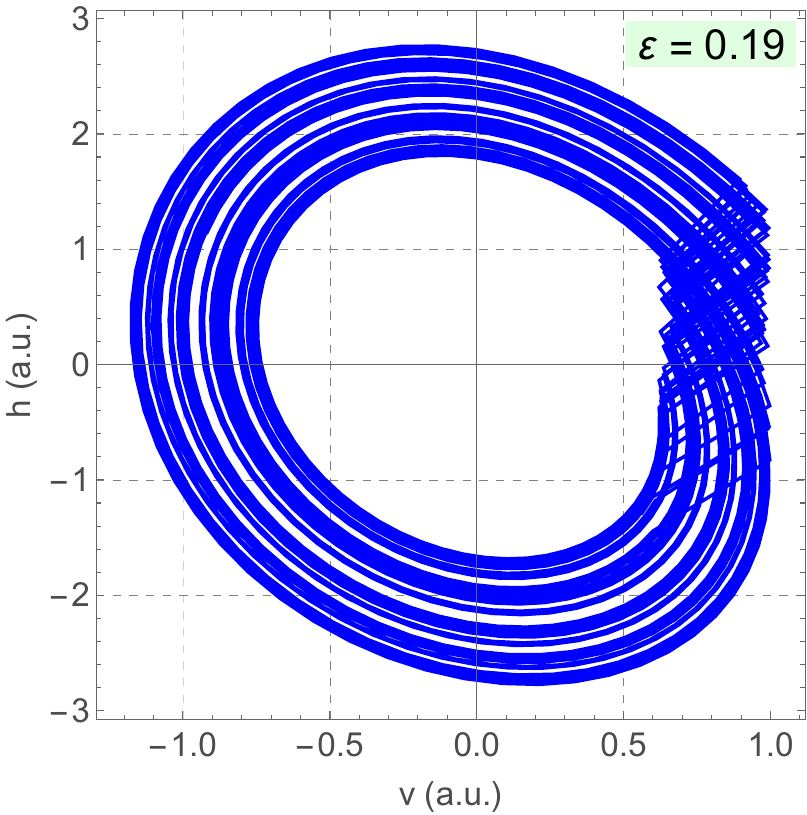}
\caption{\footnotesize \footnotesize Simulations of system (\ref{impulse1})-(\ref{impulse2}) for parameters (\ref{param}), $v_{th}=1$, and varying $\eps.$} \label{fignew12}
\end{figure*}

\vskip0.2cm

\noindent By considering a particular value 
$$
v_{th}=1,
$$
simulations of Fig.~\ref{fignew12} document that for $\eps>0$ sufficiently small the dynamics of (\ref{impulse1})-(\ref{impulse2}) is indeed a cycle of one impact per oscillation.

\vskip0.2cm

\noindent We then simulate the dynamics of (\ref{impulse1})-(\ref{impulse2}) for increasing values of $\eps>0$ to examine how violation of the assumption of Theorem~\ref{neuronthm} about the smallness of $\eps>0$ manifests itself in qualitative properties of oscillations. We conclude that increase of $\eps$ from 0 to $\eps=0.25$ first increases the number of impacts per period of the discontinuous cycle (i.e. increases the period of the cycle) and then decreases this number (see Figs.~\ref{fignew12}). We were unable to observe that increasing period is period doubling. Simulations for the values of parameters of $\eps$ between cycles of different periods suggest  chaotic windows. Indeed, the intervals of $\eps$ between the values $\eps=0.08$, $\eps=0.1,$ $\eps=0.18,$ $\eps=0.2$, $\eps=0.25$ all contain a seemingly chaotic discontinuous cycle as the one shown in Fig.~\ref{fignew12} for $\eps=0.19.$

\vskip0.2cm

\noindent We recall that according to Remark~\ref{remark1}, the left end of the interval (\ref{interv1}) is the critical value of parameter $v_{th}$ where the 1-period cycle is supposed to loose stability. Simulation of  Fig.~\ref{figunstabstab} shows perfect agreement with this prediction: 1-period cycle transforms to cycle of higher period when $v_{th}$ crosses $v_{th}=0.9156$ from the right to the left.

\vskip0.2cm

\noindent Finally, we would like to  demonstrate that parameter prediction of Theorem~\ref{neuronthm} is important for making the spikes of system (\ref{impulse1})-(\ref{impulse2}) to occur at the 
 top of membrane potential, rather than to occur on the rising phase of potential. Indeed, in Fig.~\ref{figcrest} we violate conditions of Theorem~\ref{neuronthm} in two different ways (by reducing $v_{th}$ and by increasing $\eps$) and observe that each of the two ways shifts the spikes from the top of membrane potential.

}

\begin{figure*}[!ht]\center
\vskip0.1cm
\noindent\includegraphics[scale=0.447]{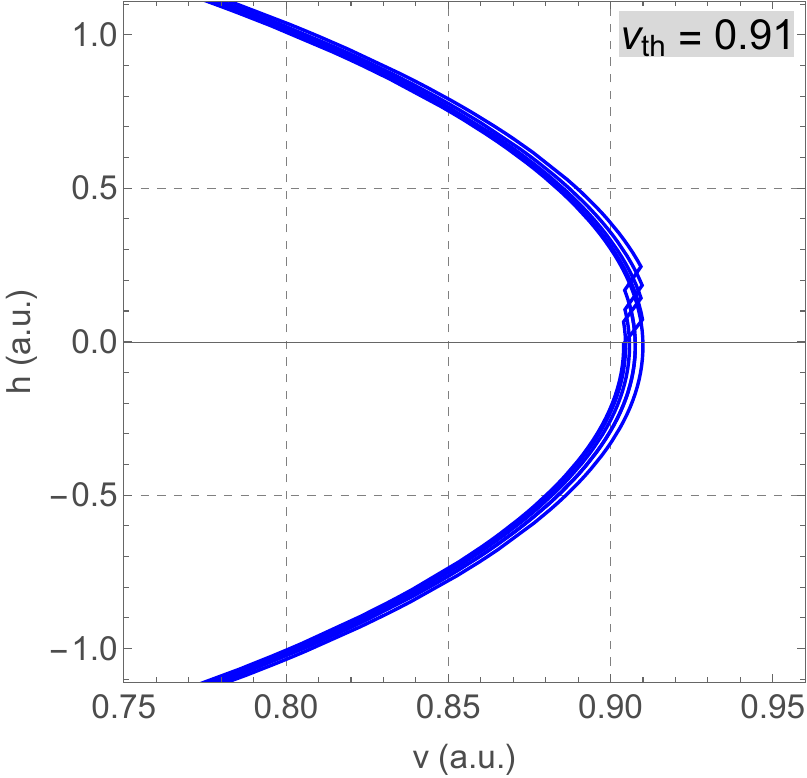}\hskip1cm
\includegraphics[scale=0.447]{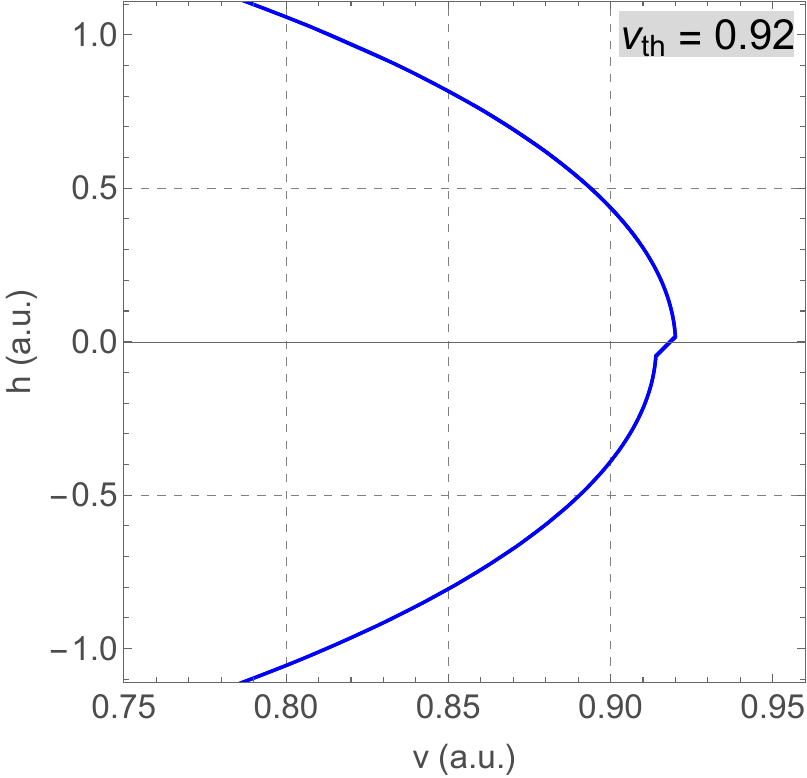}
\caption{\footnotesize The  attracting solutions of system (\ref{impulse1})-(\ref{impulse2}) with $v_{th}<0.9156$ and $v_{th}>0.9156$, $\eps=0.003$, and other parameters as given by  (\ref{param}). The solutions form limit cycle in both cases, but we zoom a part of limit cycles to make the impact visible. } \label{figunstabstab}
\end{figure*}

\begin{figure*}[!ht]\center
\vskip0.1cm

\noindent\includegraphics[scale=0.4]{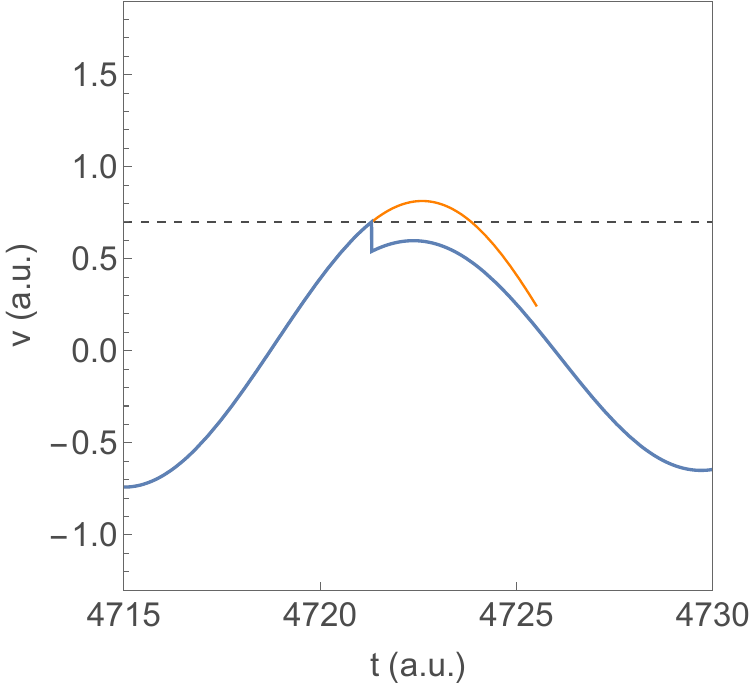}
\noindent\includegraphics[scale=0.4]{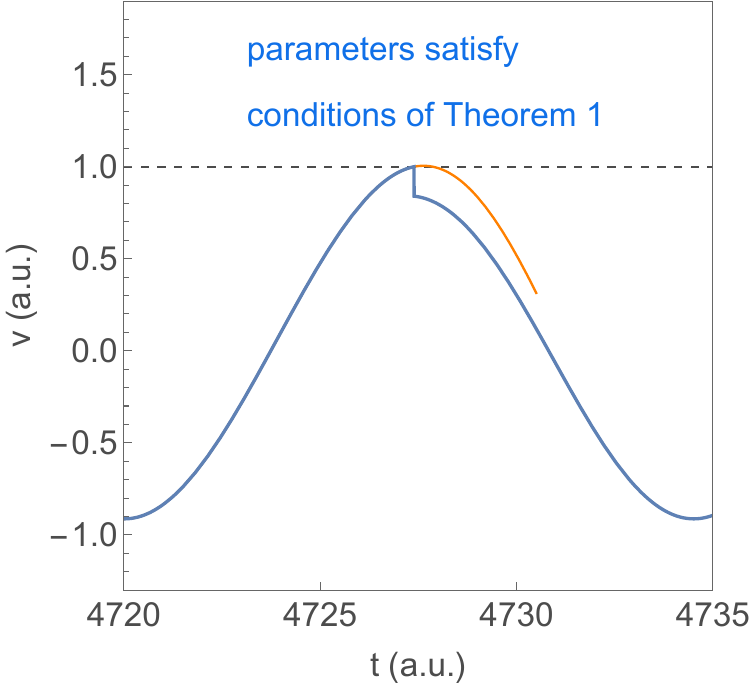}
\noindent\includegraphics[scale=0.381]{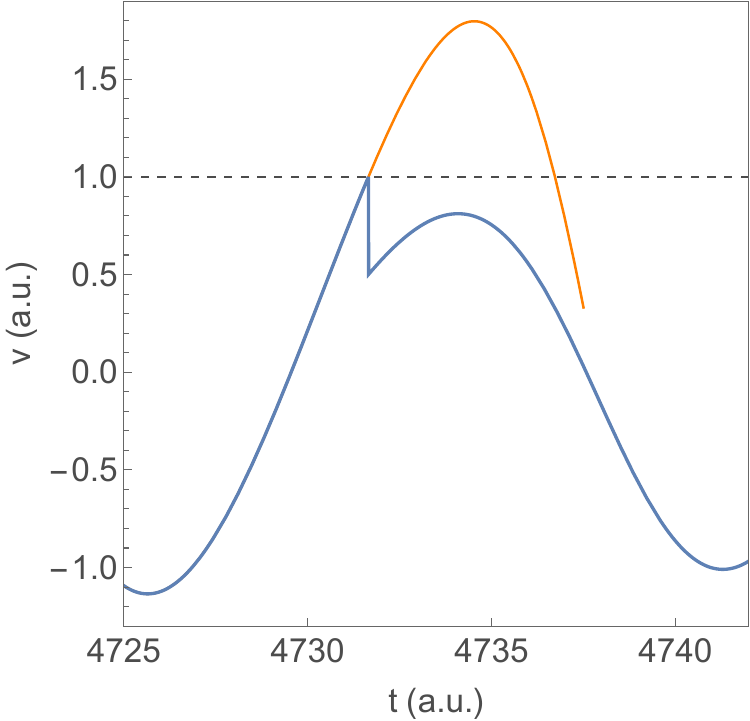}

\vskip0.2cm
\caption{\footnotesize The membrane potential $v(t)$ is shown as a function of time for the model (\ref{impulse1})-(\ref{impulse2}) with the parameters (\ref{param}) except for the following: Left: $v_{th}=0.7$, $\eps=0.08$, Middle: $v_{th}=1$, $\eps=0.08$, Right: $v_{th}=1$, $\eps=0.25$.  
} \label{figcrest}
\end{figure*}

\subsubsection{The case of stable focus}\label{sec:stablefocus}

\begin{figure}[h!]\center
\vskip0.1cm
\noindent\includegraphics[scale=0.447]{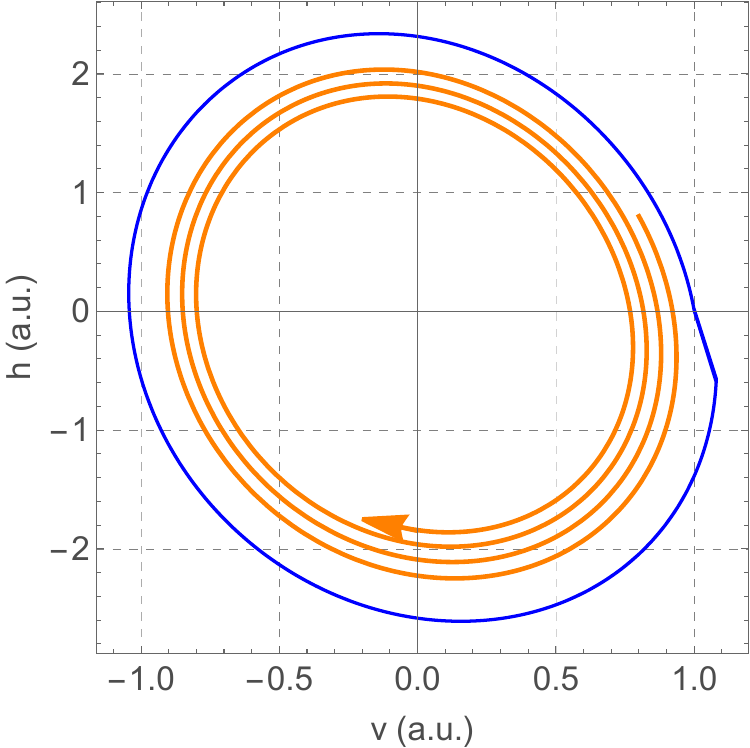}
\caption{\footnotesize The  attracting solutions of system (\ref{impulse1})-(\ref{impulse2}) with $\eps=0.08$, $k_1=0.18$, $m=0.3,$ $m_1=0.645$, $\bar h=-1$, $k_2=-1$, $m_2=0.748$, $\bar v=1$, $v_{th}=1.$ } \label{stable-focus}
\end{figure}

Because conditions of Theorem~\ref{neuronthm} concern the local behavior of system (\ref{impulse1})-(\ref{impulse2}) (near a grazing cycle of (\ref{reduced})), Theorem~\ref{neuronthm} is capable to design asymptotically stable spiking oscillations not only when the center of (\ref{reduced}) transforms to unstable focus of (\ref{impulse1}), but also in the less intuitive case when the center of (\ref{reduced})
transforms to an asymptotically stable focus of (\ref{impulse1}) (see Fig.~\ref{stable-focus}), meaning that Theorem~\ref{neuronthm} is capable to design multi-stability.

\subsubsection{The case of repeated impacts}
\begin{figure*}[!ht]\center

\vskip0.1cm
\noindent\includegraphics[scale=0.4]{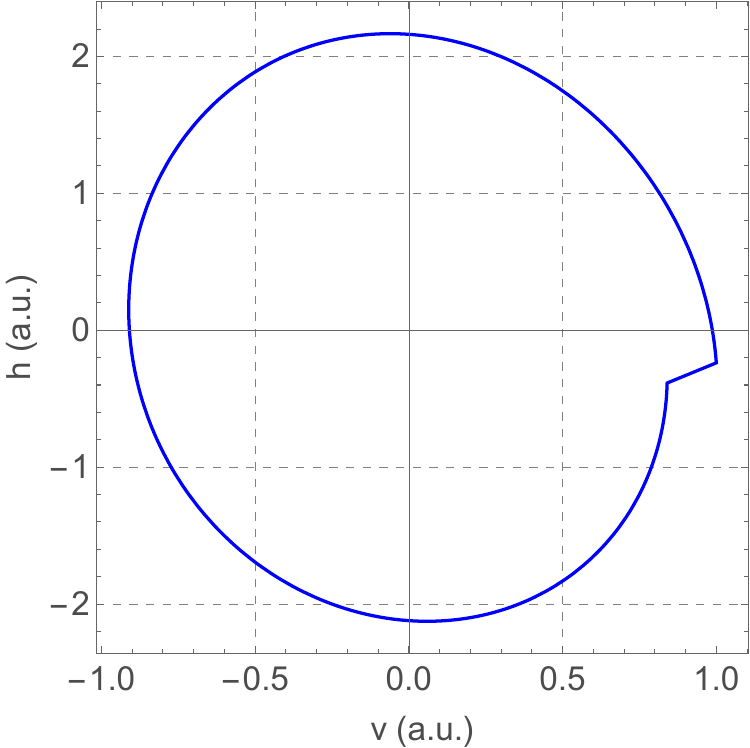}\hskip0.3cm
\noindent\includegraphics[scale=0.4]{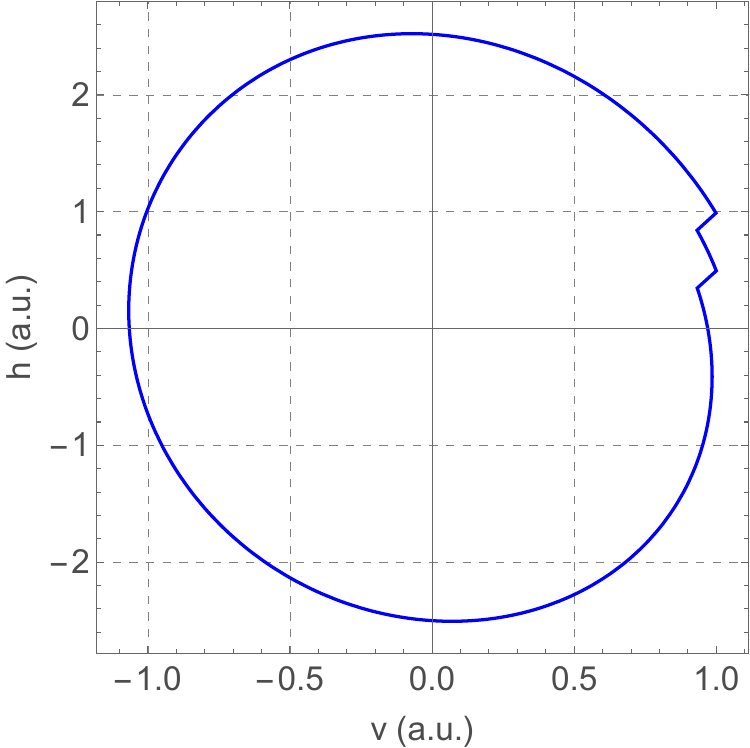}\hskip0.3cm
\noindent\includegraphics[scale=0.4]{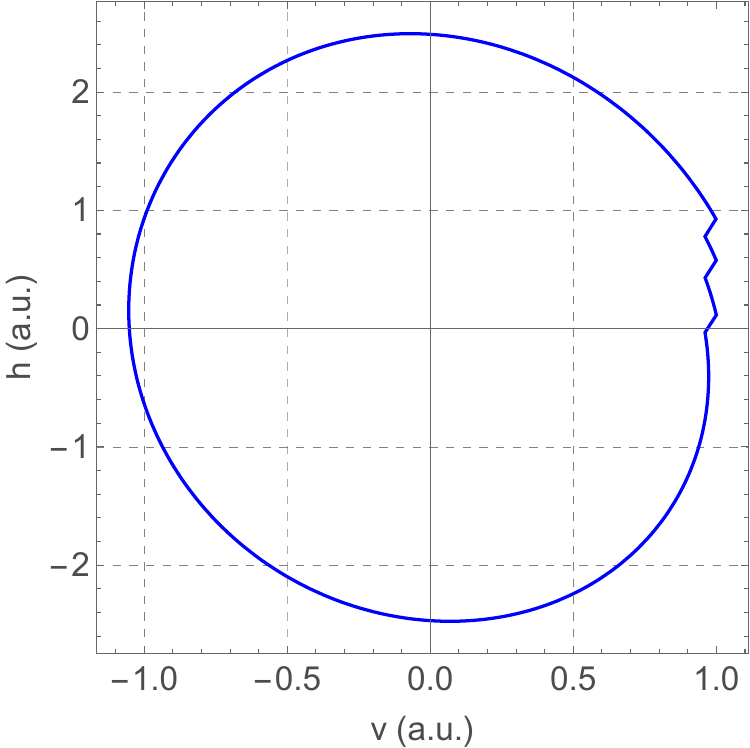}
\caption{\footnotesize The  attracting solutions of system (\ref{impulse1})-(\ref{impulse2}) with $\eps=0.08$, $k_1=0.18$, $m=0.3,$ $m_1=0.645$, $\bar h=-0.5$, $k_2=-1$, $m_2=-0.35$, $v_{th}=1.$ The value of $\bar v$ is varied: Left: $\bar v=-2$, Middle: $\bar v=-0.85$, Right: $\bar v=-0.5$.} 
 \label{fig:repeated}

\end{figure*}

Even though condition (\ref{condition2}) prevents repeated impacts in the proof of Theorem~\ref{neuronthm}, repeated impacts are possible when condition (\ref{condition2}) is violated, as Fig.~\ref{fig:repeated} documents. Moreover, Fig.~\ref{fig:repeated} suggests that any number of repeated impacts per oscillation (and close to the peak of membrane potential) is possible. Therefore, generalization of Theorem~\ref{neuronthm} for the case of a desired number of repeated impacts is a feasible project.

\subsection{Non-autonomous case}

\begin{figure}[h]\center
\vskip0.1cm
\noindent\includegraphics[scale=0.65]{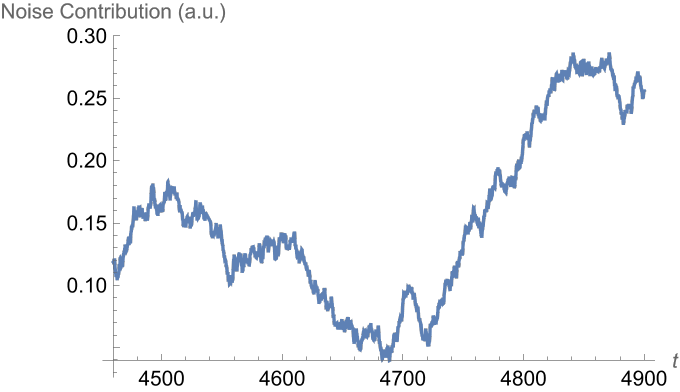}
\vskip-0.0cm
\caption{\footnotesize The graph of Wiener process $noise(t)$ as defined in (\ref{eqnoise}).} \label{fig:noise}
\end{figure}

\begin{figure}[h]\center
\vskip0.1cm
\noindent\includegraphics[scale=0.47]{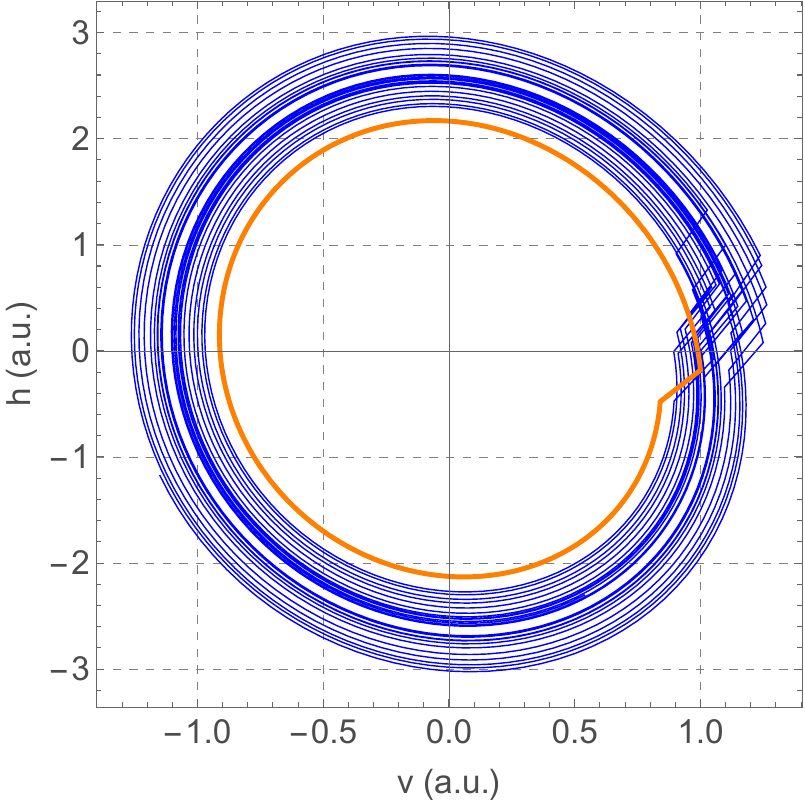}
\vskip-0.0cm
\caption{\footnotesize The limit cycle of model (\ref{impulse1})-(\ref{impulse2}) (the closed orange curve) with parameters given by  (\ref{param}), and $v_{th}=1$, $\eps=0.08$ and the corresponding solution of the system of (\ref{impulse1}) and (\ref{impulse2p}) (the black trajectory with multiple resets). The black (outer) trajectory cycles in a clockwise direction, resetting when it crosses the (moving) threshold.} \label{fig3}
\end{figure}

\noindent In this section we analyze numeric simulations of the model of (\ref{impulse1}) with the reset rule (\ref{impulse2p}) and with particular parameters of the Wiener process $noise(t)$ of (\ref{impulse2p}). 
Specifically, in this section,
\begin{equation}\label{eqnoise}
\begin{array}{l}
noise(t)\mbox{ is a Wiener process with drift }0,\\
\mbox{volatility }0.5,\mbox{ and scaled by a factor }0.01. 
\end{array}
\end{equation}The corresponding Wolfram Mathematica code reads as
\begin{itemize}
    \item[] SeedRandom[seed2use];\\
sample = RandomFunction[WienerProcess[0, 0.5], $\{$0, timebound, 0.01$\}$];\\
data = TimeSeries[sample, 
   ResamplingMethod -$>$ {"Interpolation", InterpolationOrder -$>$ 1}];\\
noise[t\textunderscore ] = 0.01*data["PathFunction"][t]; 
\end{itemize}
The graph of $noise(t)$ is shown in Fig.~\ref{fig:noise} and the location of oscillations of the system of (\ref{impulse1}) and (\ref{impulse2p}) relative to the location of oscillations of system (\ref{impulse1})-(\ref{impulse2}) is shown in Fig.~\ref{fig3}.

\vskip0.2cm

\begin{figure*}[!ht]\center
\vskip0.1cm
\noindent\includegraphics[scale=0.5]{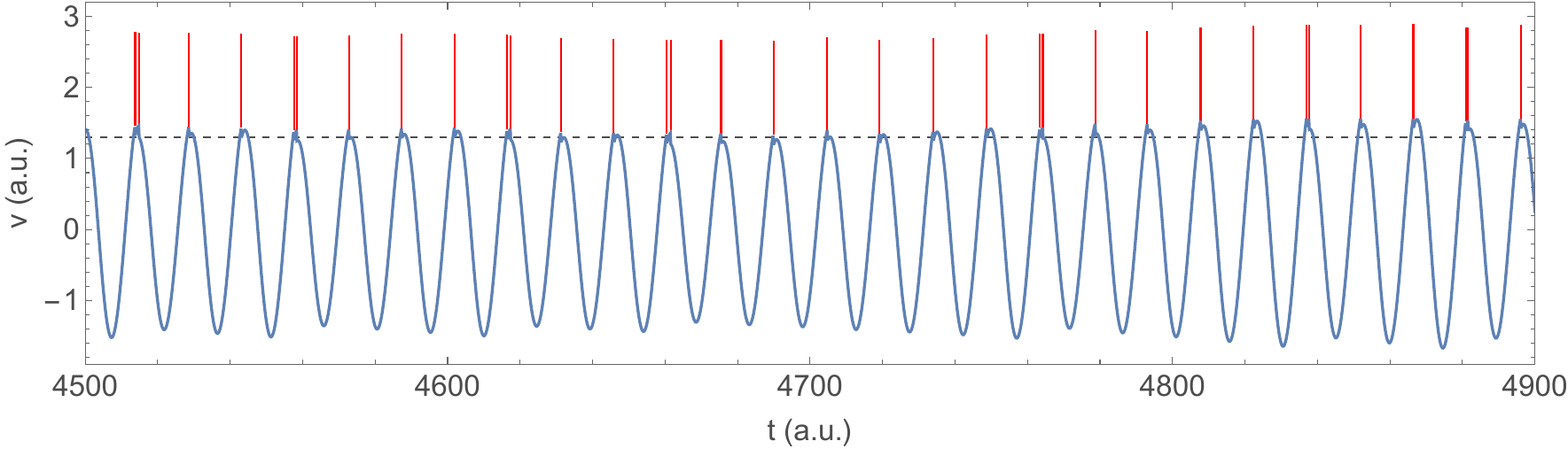}

\vskip0.25cm

\noindent\includegraphics[scale=0.5]{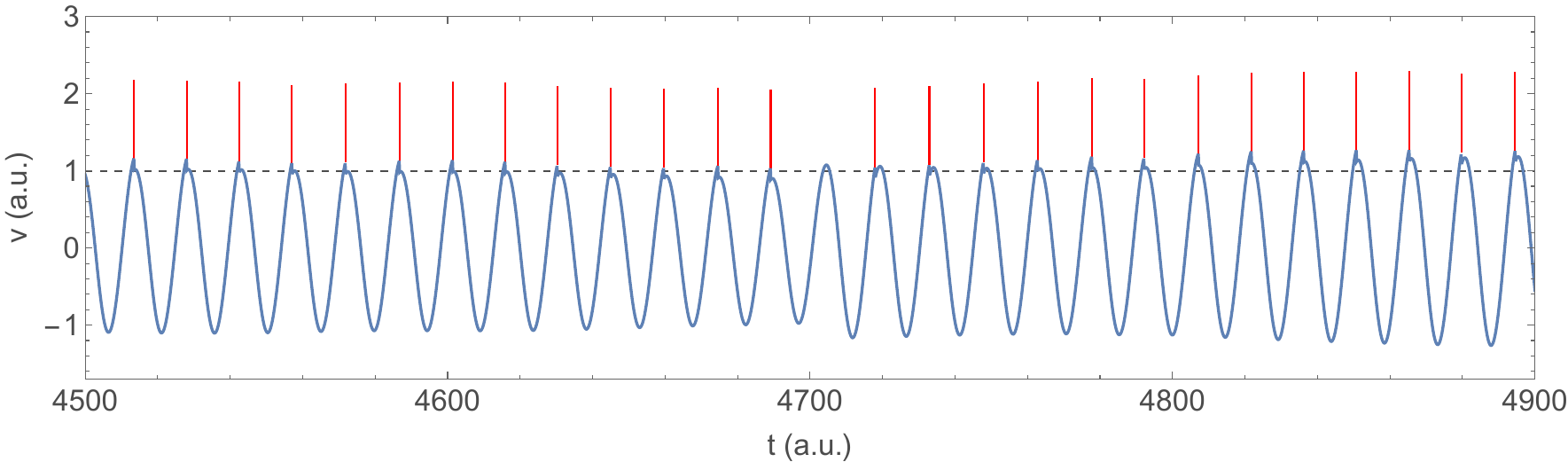}

\vskip0.3cm

\noindent\includegraphics[scale=0.5]{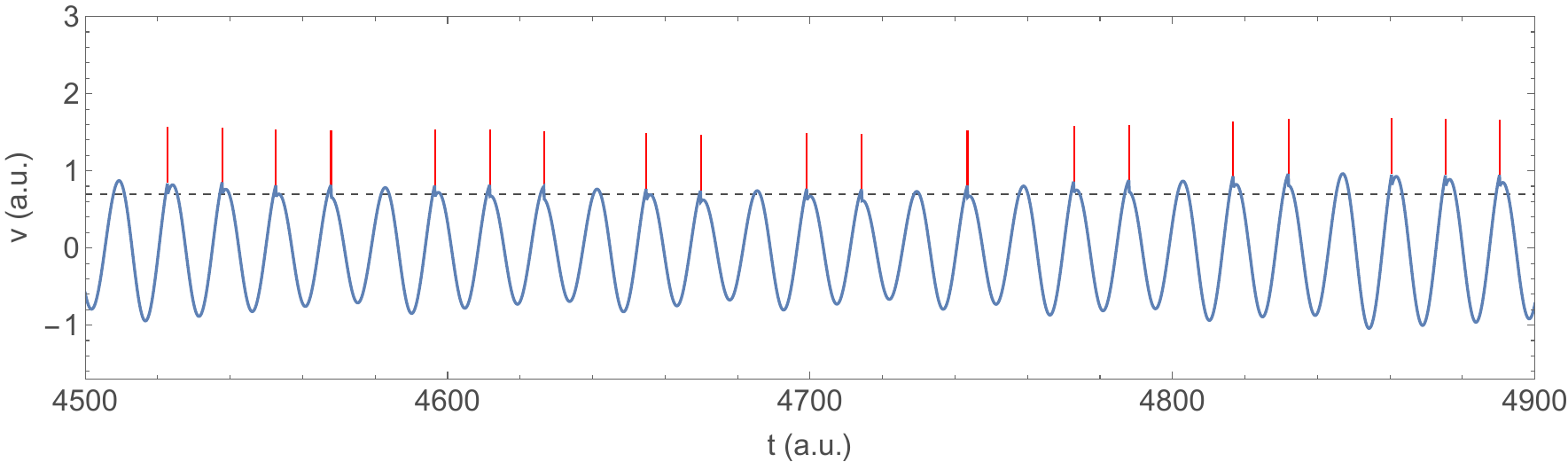}

\vskip0.2cm
\caption{\footnotesize The membrane potential $v(t)$ is shown as a function of time for the model (\ref{impulse1}) with noise added to the reset threshold according to (\ref{impulse2p}). The parameters are set as (\ref{param}) and (\ref{eqnoise}). The value of $\eps$ is $\eps=0.08$ and  $v_{th}$ is taken as $v_{th}=1.3$ (top figure), $v_{th}=1$ (middle figure), $v_{th}=0.7$ (bottom figure). The values of $v_{th}$ is represented by the horizontal dashed lines. Note, the actual reset value is not $v_{th}$, but $v_{th}$ with noise added according to (\ref{impulse2p}) and (\ref{eqnoise}).  The spikes were added manually at each reset (in red). 
} \label{fig4}
\end{figure*}

\noindent From simulations of Fig.~\ref{fig4} we conclude that parameters' prediction of Theorem~\ref{neuronthm} are remarkably robust. Indeed, noise perturbation has almost no influence on the dynamics predicted by Theorem~\ref{neuronthm}  when parameters of the unperturbed system satisfy the conditions of the theorem (i.e. we do see exactly 1 spike per oscillation in the middle simulation of Fig.~\ref{fig4}). Violation of conditions of Theorem~\ref{neuronthm} for the unperturbed system by varying $v_{th}$ beyond the requirement~(\ref{interv1}) leads to either smaller number of spikes per oscillation (bottom simulation of Fig.~\ref{fig4}) or  greater number of spikes per oscillation (top simulation of Fig.~\ref{fig4}), but the system keeps spiking in either case because the presence of unstable focus keeps widen the trajectory until it reaches the reset threshold.  

\vskip0.2cm

\noindent Overall, we conclude that the graph of $v$-dynamics of the system of (\ref{impulse1}) and (\ref{impulse2p}) (Fig.~\ref{fig4}) resembles some typical aspects of the dynamics of Fig.~\ref{nature-fig} of \citet{Domnisoru2013}.

\vskip0.2cm

\begin{figure*}[!ht]\center
\vskip0.1cm

\noindent\includegraphics[scale=0.5]{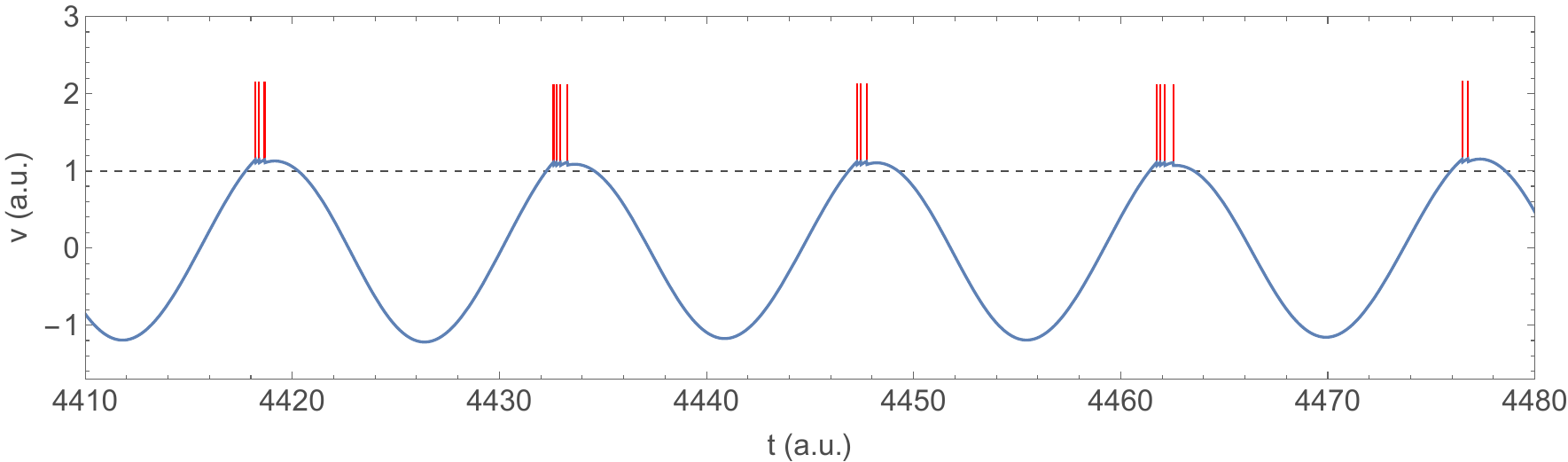}

\vskip0.2cm
\caption{\footnotesize The membrane potential $v(t)$ as in Fig.~\ref{fig4} except for parameters changed as follows: $\bar h=-0.5$, $\bar v=-0.5$, $v_{th}=1.$
} \label{fig4new}
\end{figure*}

\noindent Additionally, in Fig.~\ref{fig4new} we simulate the system of (\ref{impulse1}) and (\ref{impulse2p}) for the parameters corresponding to the case of the repeated spiking of Fig.~\ref{fig:repeated}(right) to demonstrate that the qualitative properties of the dynamics of Fig.~\ref{fig:repeated} (repeated spiking close to the peak of membrane potential) are robust with respect to noise perturbation, meaning that developing an analogue of Theorem~\ref{neuronthm} for the case of repeated spiking promises to be a successful project that will provide further analytic support to the dynamics of Fig.~\ref{nature-fig} of \citet{Domnisoru2013}.

\section{Discussion}
The model presented here demonstrates how individual neurons could maintain spiking that is phase locked to the peak or right before the peak of intracellular membrane potential oscillations.  An important component of this process is a model with an unstable center, resulting in a neuron that shows progressively larger oscillations that spiral out from the center until reaching the spiking threshold. The reset of membrane potential after each spiking event results in a limit cycle that continues to graze the threshold on subsequent cycles, resulting in spikes that occur near the peak of a cycle. Biologically, an unstable center could be provided by intrinsic ionic mechanisms such as those that support spontaneous subthreshold theta oscillations \citep{Bal1997, Boehmer2000}. 

As shown in Figures~\ref{fig3} and \ref{fig4}, this model provides a potential mechanism for the experimental observation of spikes occurring near the peak of each oscillation of membrane potential \citep{Harvey2009, Domnisoru2013, SchmidtHieber2013}. Incorporating a variable threshold as seen biologically  \citep{Higgs2011, Platkiewicz2011, Tsuno2013, Wester2013, Fontaine2014}  allows the model to replicate the data in which the membrane potential and spiking onset vary across oscillation cycles, resulting in the appearance of spikes at different membrane potentials \citep{Domnisoru2013}.

This mechanism for maintaining spiking at the peak of each oscillation could be relevant to oscillatory interference models of theta phase precession that require spikes to occur near the peak of each cycle, as required in models of theta phase precession which utilize an intrinsic oscillation that maintains intrinsic spiking at one phase relative to the network oscillation \citep{OKeefe1993, Lengyel2003, Burgess2007, Bush2014, Hasselmo2014,Hasselmo2014Phil}.  It should be noted that oscillatory interference is only one potential model of theta phase precession, as precession has also been attributed to different mechanisms including read out of sequences or the influence of an asymmetrical ramp of depolarization that interacts with oscillatory input to cause spikes to advance to earlier phases with greater depolarization \citep{Kamondi1998, Mehta2002}.

The maintenance of spiking at a specific phase of intracellular oscillations, coupled with the shift in phase of oscillations relative to network field potential could contribute to the encoding and retrieval of memories. Different phases of hippocampal theta show different functional dynamics that are proposed to underlie separate phases of encoding and retrieval \citep{Hasselmo2002}. The role of different phases is supported by neurons showing differential phase of firing for novel stimuli versus familiar stimuli \citep{Zilli2006, Manns2007, Lever2010}, by EEG traces in human performing memory tasks \citep{Rizzuto2006, Kerren2018}, and by the correlation between theta phase and eye fixations on novel versus retrieved locations \citep{Kragel2020}.  Additionally, the inhibition of hippocampal neurons causes selective effects on behavior depending on the phase of the manipulation \citep{Siegle2014}. 

To produce this neuron model that maintains a spiking phase preference relative to the intracellular theta oscillation, we began with a simple linear model that resembles the nonlinear Izhikevich neuron model for resonate-and-fire neurons with some key differences.  The inclusion of a small parameter $\epsilon$ introduced an unstable focus to the system. The unstable focus of our model cell differs from Izhikevich models, which generally employ a stable resting potential or an oscillating potential that dampens with time. However, there is experimental evidence for sustained subthreshold oscillations and other spontaneous intrinsic activity in vivo \citep{Bal1997, Boehmer2000, Hausser2004}. 

Additionally, the spiking threshold of our model varied over time. Variability in spike threshold, as seen biologically \citep{Higgs2011, Platkiewicz2011, Tsuno2013, Wester2013, Fontaine2014}, is not often implemented in spiking cell models. However, the unstable focus in the membrane potential of our model enabled the cell to reliably reach the variable spiking threshold.  Biologically, the parameter $\epsilon$ may be interpreted as a contribution of multiple ion channel types to the neuron. The membrane potential dynamics are likely to incorporate both an afterhyperpolarizing potassium current (in the reset equation) and a depolarizing mixed cation current (in the voltage equation).

\message{LaTeX Warning: Oleg may add some details about how this work relates to previous work, such as a reference to the paper using the v bar term \the\inputlineno}

Models of the cellular contributions to spatial processing in brain networks continue to play a significant role in advancing our ideas of how the brain produces spatial cognition. The equations of our model cell are straightforward to implement and have the benefit of modeling a cell that can fire near the peak of the theta rhythm even on theta cycles where the baseline potential was relatively depolarized, similar to dynamics observed in vivo \citep{Domnisoru2013}. We hope these equations can be of use to modelers and others in the study of spatial processing.

\section*{Acknowledgments} This collaboration was made possible by the Burroughs Wellcome Fund with a
Collaborative Research Travel Grant \#1017453 to OM to visit MH at the Center for Systems Neuroscience
at Boston University. OM thanks the Center for providing excellent working conditions. This work was also
supported by the National Institutes of Health and the Office of Naval Research. The authors are grateful to anonymous reviewers for useful comments that helped to improve the quality of the paper.

\section*{Declarations}
\begin{itemize}
\item Conflict of interest/Competing interests. The author declares no competing interests.
\end{itemize}

\begin{appendices}
\section{Proofs of lemmas}    

\noindent {\bf Proof of Lemma~\ref{lemsingular}.} By expressing $(\eps,h^2)$ in polar coordinates (for $r>0,$ $\phi\in[0,2\pi]/\{0,2\pi\}$),
$$
   \eps=r\sin\phi,\quad h^2{\rm sign}(h)=r\cos\phi,
$$
i.e.  $0$ and $2\pi$ are identified, 
we observe that equation (\ref{genform}) can be rewritten as
$$
\begin{array}{l}
   t^2 + at\sqrt{r|\cos\phi|}\cdot\alpha+br\sin\phi +ct r\sin\phi+\\
   +p\sqrt{r|\cos\phi|}r\sin\phi\cdot\alpha+qr^2\sin^2\phi+\\
  \hfill +I(t,r,\phi)\left(\begin{array}{c} t\\ \sqrt{r|\cos\phi|}\cdot\alpha\\ r\sin\phi\end{array}\right)^3=0,\quad \alpha={\rm sign}(h)
   \end{array}
$$
which contains just one small parameter $r>0$ as opposed to the two small parameters $\eps>0$ and $h>0$ in the initial equation (\ref{genform}). Moreover, the equation obtained suggests that $t\sim\sqrt{r}$, which hints at searching for $t$ in the form
$$
   t=\sqrt{r}s.
$$
This change of the variable yields, upon dividing by $r$,
\begin{equation}\label{quadr}
\begin{array}{l}   s^2+as\sqrt{|\cos\phi|}\cdot\alpha+b\sin\phi+c\sqrt{r}s\sin\phi+\\
+p\sqrt{r|\cos\phi|}\sin\phi\cdot\alpha+qr\sin\phi+J(s,r,\phi)=0,
\end{array}
\end{equation}
where
$$
\begin{array}{l}
  J(s,r,\phi)=\dfrac{I(s,r,\phi)}{r}\left(\begin{array}{c} \sqrt{r}s\\ \sqrt{r|\cos\phi|}\cdot\alpha \\ r\sin\phi\end{array}\right)^3=\\
  \sqrt{r} I(s,r,\phi)\left(\begin{array}{c} s\\ \sqrt{|\cos\phi|}\cdot\alpha \\ \sqrt{r}\sin\phi\end{array}\right)^3.
\end{array}
$$
Denoting the left hand side of (\ref{quadr}) by $F(s,\sqrt{r},\phi)$, we observe that the two solutions $s$ of $F(s,0,\phi)=0$ are given by
$$
   s_\pm(\phi)=\dfrac{1}{2}\left(-a\sqrt{|\cos\phi|}\cdot\alpha\pm\sqrt{a^2|\cos\phi|-4b\sin\phi}\right),
$$
provided that $a^2|\cos\phi|-4b\sin\phi>0.$ Moreover, $F_s(s_\pm(\phi),0,\phi)=2s_\pm(\phi)+a\sqrt{|\cos\phi|}\cdot\alpha=\pm\sqrt{a^2|\cos\phi|-4b\sin\phi}\ {\color{black}=\pm\dfrac{(ah)^2-4b\eps}{\sqrt{h^4+\eps^2}}.}$ 
Therefore, by the implicit function theorem {\color{black}(that applies thanks to condition (\ref{separated}))}, given $\gamma>0$ there exists $\delta>0$ such that 
for any $r\in[0,\delta]$ and for any $\phi\in[0,2\pi]/\{0,2\pi\}$ such that $a^2|\cos\phi|-4b\sin\phi\ge \gamma$, the equation  $F(s,\sqrt{r},\phi)=0$ has two solutions: a unique solution $S(\sqrt{r},\phi)\to s_{-}(\phi)$ as $r\to 0$ {\color{black}and a unique solution  
approaching $s_+(\phi)$ as $r\to\infty$. Since $s_-(\phi)<s_+(\phi)$ (i.e. $s_-(\phi)$ is the least solution) it remains to show that $S(\sqrt{r},\phi)$ matches the formula that the lemma claims.} 
By the implicit function theorem, the solution $(\sqrt{r},\phi)\to S(\sqrt{r},\phi)$ is $C^2$ in the above-defined domain of definition, see e.g. \cite[\S8.5.4, Theorem~1]{Zorich}. Therefore $S(\sqrt{r},\phi)$ can be expanded in Taylor series as follows
$$
\begin{array}{l}   S(\sqrt{r},\phi)=S(0,\phi)+I(\sqrt{r},\phi)\sqrt{r}=\\ =\dfrac{1}{2}\left(-a\sqrt{|\cos\phi|}\cdot\alpha-\sqrt{a^2|\cos\phi|-4b\sin\phi}\right)+\\+I(\sqrt{r},\phi)\sqrt{r},
\end{array}
$$
where $I$ is $C^1$ smooth on $r\in[0,\delta]$, $\phi\in[0,2\pi]/\{0,2\pi\},$ and $I_\phi(0,0^+)=I_\phi(0,2\pi^-).$
Returning back to the variable $t$, we get
\begin{equation}\label{t}
\begin{array}{l}  t=\sqrt{r}S(\sqrt{r},\phi)=\dfrac{1}{2}\left(-ah-\sqrt{(ah)^2-4b\eps}\right)+\\
+I(h,\eps)\sqrt{h^4+\eps^2},
\end{array}
\end{equation}
where $I$ is $C^1$-smooth on $\sqrt{h^4+\eps^2}\le \delta.$ 
The proof of the lemma is complete.

\vskip0.2cm

\noindent {\bf Proof of Lemma~\ref{lemma2}.} By solving the linear system  (\ref{reduced}), one gets
\begin{eqnarray}
&&  \left(\begin{array}{c}
    V(t,v,h,0)\\
    H(t,v,h,0)\end{array}\right)
=\Omega(t)\left(\begin{array}{c}
    v\\
    h\end{array}\right),\label{VHformula}\\ 
&& \Omega(t)= \left(\begin{array}{cc}
    \cos(\omega t) & (k_1/\omega)\sin(\omega t)\\
    (k_2/\omega)\sin(\omega t) & \cos(\omega t)\end{array}\right).\nonumber
\end{eqnarray}
{\color{black}In particular, $\Omega(T_0)=I$, $\Omega'(T_0)=\left(\begin{array}{cc} 0 & k_1\\ k_2 & 0\end{array}\right),$ $\Omega''(T_0)=\left(\begin{array}{cc} -\omega^2 & 0\\ 0 & -\omega\end{array}\right),$ $V(x_0)=v_{th},$ $H(x_0)=0$.
Thus, (\ref{VHformula}) implies 
$$
\begin{array}{rcl}
&&V_{v}(x_0)=H_h(x_0)=1,\ \ V_h(x_0)=H_v(x_0)=0,\\ 
&&  V_t(x_0)=k_1 H(x_0)=0,\\
&&H_t(x_0)=k_2V(x_0)=k_2v_{th},\\
&&V_{th}(x_0)=k_1,\ V_{tt}(x_0)=-\omega^2 v_{th}.
\end{array}
$$

\noindent Since $V(t,v,h,0)$ and $H(t,v,h,0)$ are linear in $(v,h),$ we get that all second-order partial derivatives of $V$ and $H$ with respect to $v$ and $h$ (including second-order mixed partial derivatives) equal 0.}

\vskip0.2cm

\noindent To compute the derivative $(V_\eps,H_\eps)$ we differentiate (\ref{impulse1}) with respect to $\eps$ obtaining a linear inhomogeneous system 
\begin{equation}\label{Ay}
   \dot y=\left(\begin{array}{cc} 0 & k_1\\ k_2 & 0 \end{array}\right)y+f(t), 
\end{equation}
where 
$$
\begin{array}{l}
 y=\left(\begin{array}{c}
    V_{\eps}(t,v_{th},0,0)\\
    H_{\eps}(t,v_{th},0,0)\end{array}\right), \\  f(t)=\left(\begin{array}{c}
    mV(t,v_{th},0,0)^2+m_1 V(t,v_{th},0,0)\\
    m_2 H(t,v_{th},0,0)\end{array}\right).
\end{array}
$$
In particular, we have $(V_\eps(x_0),H_\eps(x_0))^T=y(T_0)$.
Noticing that $y(0)=0$, 
the variation of constants formula yields
\begin{equation}\label{vari}
y(t)
=\Omega(t)\int_0^t\Omega(-s)f(s)ds,
\end{equation}
{\color{black}whose full computation is provided in formula (\ref{qq}).
\begin{figure*}[!ht]
{\color{black}
\begin{equation}\label{qq}
y(t)=\left(\begin{array}{c}
\cos(\omega t)\left(\dfrac{2\,m\, v_{th}^2\sin(\omega t)}{3\omega}+\dfrac{tv_{th}(m_1+m_2)}{2}\right)+v_{th}\sin(t\omega)\dfrac{2\,m\,v_{th}+3(m_1-m_2)}{6\omega}\\
\dfrac{m\,v_{th}^2\cos(\omega t)}{3k_1}-\dfrac{m\,v_{th}^2\sin^2(\omega t)}{3k_1}-\dfrac{t\,v_{th}(m_1+m_2)\omega\sin(\omega t)}{2k_1}-\dfrac{m\,v_{th}^2}{3k_1}
\end{array}\right)
\end{equation}
}
\end{figure*}}
The required formula for $(V_\eps(x_0),H_\eps(x_0))$ follows by plugging $t=T_0=2\pi/\omega$ in  formula (\ref{qq}).

\vskip0.2cm

\noindent The proof of the lemma is complete.

\end{appendices}

\bibliography{grazingpaperBIB}

\end{document}